\documentclass[aps,prb,twocolumn,floatfix]{revtex4-1}
\usepackage{times}
\usepackage{epsfig} 
\usepackage{amsmath}
\usepackage{amssymb} 
\usepackage{graphicx}
\usepackage{dsfont}
\usepackage{bbm}
\usepackage[colorlinks=true,citecolor=red,linkcolor=blue]{hyperref}

\newcommand{\bfk}{\mathbf{k}}
\newcommand{\bfa}{\mathbf{a}}
\newcommand{\bfb}{\mathbf{b}}
\newcommand{\bfA}{\mathbf{A}}
\newcommand{\bfR}{\mathbf{R}}

\newcommand{\mathJ}{\mathcal{J}}
\newcommand{\bfK}{\mathbf{K}}
\newcommand{\bfM}{\mathbf{M}}

\newcommand{\GKM}{\mathrm{GKM}}
\newcommand{\DKM}{\mathrm{DKM}}

\begin{document}

\title{Floquet topological transitions in extended Kane-Mele models with disorder}
\author{Liang Du}
\author{Paul D. Schnase}
\author{Aaron D. Barr}
\author{Ariel R. Barr}
\author{Gregory A. Fiete}
\affiliation{Department of Physics, The University of Texas at Austin, Austin, TX 78712, USA}

\begin{abstract}
In this work we use Floquet theory to theoretically study the influence of circularly polarized light on disordered two-dimensional models exhibiting topological transitions.   We find circularly polarized light can induce a topological transition in extended Kane-Mele models that include additional hopping terms and on-site disorder. The topological transitions are understood from the Floquet-Bloch band structure of the clean system at high symmetry points in the first Brillouin zone.  The light modifies the equilibrium band structure of the clean system in such a way that the smallest gap in the Brillouin zone can be shifted from the $M$ points to the $K(K')$ points, the $\Gamma$ point, or even other lower symmetry points.  The movement of the minimal gap point through the Brillouin zone as a function of laser parameters is explained in the high frequency regime through the Magnus expansion.  In the disordered model, we compute the Bott index to reveal topological phases and transitions. The disorder can induce transitions from topologically non-trivial states to trivial states or vice versa, both examples of Floquet topological Anderson transitions.  As a result of the movement of the minimal gap point through the Brillouin zone as a function of laser parameters, the nature of the topological phases and transitions is laser-parameter dependent--a contrasting behavior to the Kane-Mele model.
\end{abstract}
\date{\today}
\maketitle

\section{INTRODUCTION}
\label{sec:intro}
Research on topological band insulators has seen dramatic progress in the past decade.\cite{Moore:nat10,Hasan:rmp10,QiX:rmp11,Ando:jpsj13}  The phenomenology is even richer when inter-particle interactions are taken into account and fractionalized phases result. \cite{Maciejko:np15,Stern:arcmp15,witczak-krempa:arcmp14,mesaros:prb13,ChenX:prb13}  Starting from a non-interacting band structure, the Coulomb interaction can induce a topological transition.\cite{SunK:prl2009,XueF:prb17,XueF:arXiv17}  For example, in the two-dimensional honeycomb lattice, the Dirac points are stable to weak Coulomb interaction, while the bulk gap will open at a finite critical Coulomb interaction.\cite{Meng:nat10,Hohenadler:prl11,YuS:prl11} In the kagome lattice, there is a flat band and a quadratic band touching point which is perturbatively unstable to the Coulomb interaction.\cite{SunK:prl2009,WangY:prl11}  Recently, an active direction of research has been to study the topological transition by periodically driving a non-interacting system to a non-equilibrium state, called a Floquet topological insulator.\cite{Lindner:np11} 
A periodic drive can be realized in a cold atom system with an optical lattice potential generated by changing the laser field,\cite{Jotzu:nat14,Bilitewski:pra15} or in the solid state by illumination with a monochromatic laser field. \cite{Oka:prb09,Kitagawa:prb10,Fregoso:prb13,Sentef:nc15,WangY:sci13,Mahmood:np16,Calvo:prb15,Lago:pra15,Perez-Piskunow:pra15,Perez-Piskunow:prb14,Dehghani:prb14,Dehghani:prb15a,DAlessio:nc15,Sacksteder:prb16,DuL:prb17a,GeY:pra17,ChenQ:prb18}

In equilibrium, topological insulators induced by Anderson (on-site) disorder have been well studied in the past decade. \cite{LiJ:prl09,Groth:prl09,JiangH:prb09,JiangH:prl09,Prodan:prl10,Victor:prb11,XuD:prb12,WuQ:prb13,SongJ:prb14a,SongJ:prb14b,Girschik:prb15,Orth:sr16}
Within the Born approximation, Anderson disorder will induce a negative correction to the mass and chemical potential, which in turn may induce a topological transition.\cite{Groth:prl09} 
Song {\it et al.}\cite{SongJ:prb12} studied the effect of different types of disorder on the topological transition in the Haldane model where a Dirac point is situated at the $K,K'$ points. Their study shows that on-site disorder and bond disorder have different effects on the topological transition. Bond disorder tends to prohibit the system from undergoing a phase transition to a topological Anderson insulator, contrary to the effect of Anderson disorder. When the Kane-Mele model\cite{Kane:prl05a,Kane:prl05b} is generalized to include third-neighbor hopping, or dimerized first-neighbor hopping terms along the $z$ direction, the linear crossing can shift from a $K,K'$ point to an $M$ point.\cite{Hung:prb16}
At the $M$ point, the bond and on-site disorder have the same effect on the mass renormalization, and both enhance the topological state in the weak disorder limit.\cite{Hung:prb16}  Hung {\it et al.}\cite{Hung:prb16} studied the generalized Kane-Mele (GKM) model and dimerized Kane-Mele (DKM) model (described in this paper in Sec.~\ref{sec:model}).  They found that low and intermediate levels of disorder tend to stabilize the topological phase for both models. Further, taking the Coulomb interaction into account tends to destabilize the topological phase in the dimerized Kane-Mele model, but stabilize the topological phase in the GKM model.  Hence the GKM and DKM provide contrasting behavior to each other, and also to the more heavily studied Kane-Mele model, thus illustrating the phenomenological richness of topological phases and transitions under different conditions.

To summarize, the location of the Dirac point in momentum space in a clean (disorder-free) system is crucial to determining the effect of bond or on-site disorder.  In this paper, we show that starting from  a fixed equilibrium model Hamiltonian, periodically driving the system out-of-equilibrium via a laser can shift the Dirac point between different high symmetry points, for example, from an $M$ to a $K$ or a $\Gamma$ point. These shifts are computed in detail, and provide a platform to study differences in the effects of bond and on-site disorder in the presence of a laser field.  Out-of-equilibrium, a disorder-induced transition between topologically trivial and nontrivial states is characterized by the disorder-averaged Bott index.\cite{Loring:epl10} Prior non-equilibrium work studied the honeycomb lattice with staggered on-site A-B sub-lattice potentials in the presence of disorder.\cite{Titum:prl15,Titum:prx16}

In this paper, we focus on laser- and disorder-induced topological transitions. Before turning to the disorder-induced Floquet topological phase transition in the GKM and DKM models, we first study the Floquet-Bloch band structure where a gap closing and reopening process is observed. The effect of disorder on the clean Floquet system is studied and the results qualitatively explained considering the energy scales of the system gap size and the total bandwidth.

The organization in this paper is as follows. In Sec.\ref{sec:model}, we describe the generalized Kane-Mele and dimerized Kane-Mele models.  The Floquet topological transition, the Floquet-Bloch band structure and the related low-energy theory are described in Sec.\ref{sec:clean}. In Sec.\ref{sec:disorder}, we study the topological transition in the generalized and dimerized Kane-Mele models subject to both laser illumination and on-site disorder. Finally, in Sec.\ref{sec:conclusion}, we summarize our main conclusions.
\section{Model Hamiltonian}
\label{sec:model}
\begin{figure}[h]
\includegraphics[width=1.0\linewidth]{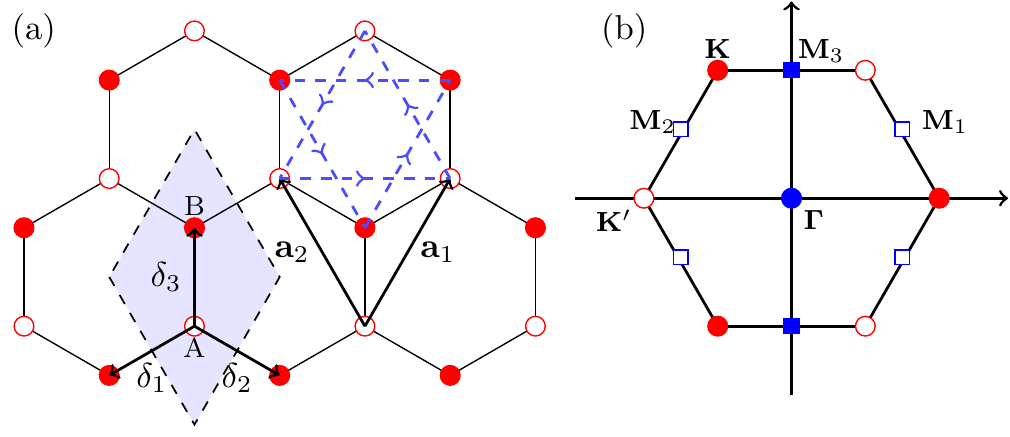}  %
\caption{(Color online) (a) Honeycomb lattice with two sub-lattices in one unit cell (shaded area), labeled A (open circles) and B (filled circles). 
Three nearest-neighbor unit vectors are $\delta_1=(-\sqrt{3}/2,-1/2)a$, $\delta_2=(\sqrt{3}/2,-1/2)a$, $\delta_3=(0,1)a$, with $a$ the nearest-neighbor distance. Lattice translational vectors are labeled as $\bfa_1=\delta_3-\delta_1=(\sqrt{3}/2,3/2)a,\bfa_2=\delta_3-\delta_2=(-\sqrt{3}/2,3/2)a$. The blue dashed lines represent the imaginary second-neighbor hopping (spin-orbit coupling) and the arrow directions represent positive signs. (b) First Brillouin zone of the underlying triangular Bravais lattice with reciprocal lattice vector $\bfb_1=(\sqrt{3},1)2\pi/(3a)$ and $\bfb_2=(-\sqrt{3},1)2\pi/(3a)$. High symmetry points are ${\bf K}=(-2\pi/\sqrt{3},2\pi)/(3a)$, ${\bf K'}=(-4\pi/\sqrt{3},0)/(3a)$ and time reversal invariant momenta ${\bf M}_{1,2}=(\pm \sqrt{3}\pi, \pi)/(3a)$, ${\bf M}_3=(0, 2\pi)/(3a)$.  All filled circles are equivalent to ${\bf K}$ and all open circles are equivalent to  ${\bf K'}$. The Floquet quasi-band structure is plotted along the momentum path $K'-\Gamma - M_3 - K - M_2 - K'$ in the first Brillouin zone.
}
\label{fig:honeycomb}
\end{figure}
We study both the generalized Kane-Mele (GKM) tight-binding Hamiltonian with third-nearest neighbor hopping terms and the dimerized Kane-Mele (DKM) model with dimerized hopping parameter in the vertical direction on the honeycomb lattice (Fig.\ref{fig:honeycomb}(a)). The GKM Hamiltonian in real-space is given by,
\begin{align}
     H_{\GKM}^\sigma =& 
     -t_1 \sum_{\langle ij \rangle} c_{i}^{\dagger} c_{j}^{}
     +i \lambda_{\text soc} \sum_{\langle\!\langle ij \rangle\!\rangle}\sigma v_{ij} c_{i}^{\dagger} c_{j} \nonumber\\
     &- t_3 \sum_{\langle\!\langle\!\langle ij \rangle\!\rangle\!\rangle}c_{i}^{\dagger} c_{j}^{}
\label{eq:htbrGKM}
\end{align}
where $t_1(t_3)$ is the isotropic hopping integral between first- (third-) nearest neighbors, 
$c_{i}^\dagger$ ($c_{j}$) creates (annihilates) an electron with spin $\sigma$ 
on site $i$ ($j$) of the honeycomb lattice (the spin subindex is omitted for simplicity), and $\langle ij \rangle$ limits the summation to nearest neighbors, $\langle\!\langle ij \rangle\!\rangle$ and $\langle\!\langle\!\langle ij \rangle\!\rangle\!\rangle$ limit the summation to second- and third-nearest neighbors, respectively. Here $\lambda_{soc}$ is the spin-orbit coupling strength, $\sigma = 1 (-1)$ for a spin-$\uparrow$ ($\downarrow$) sector Hamiltonian, $v_{ij} = 1$ for the counter-clockwise hopping shown in Fig.\ref{fig:honeycomb}(a) with dashed arrow lines, and $v_{ij} = -1$ for clockwise hopping.  In Eq.\eqref{eq:htbrGKM} only the spin-$\sigma$ part of the Hamiltonian is written explicitly. The Hamiltonian with opposite spin-$\bar{\sigma}$ is the time-reversal of $H_{\GKM}^\sigma$. 

The DKM Hamiltonian in real-space is given by,
\begin{align}\label{eq:htbrDKM}
     H_{\DKM}^{\sigma} =& 
     \sum_{i\in A} -t_1 (c_{i}^{\dagger} c_{i+\delta_1}^{} + c_{i}^{\dagger} c_{i+\delta_2}^{})
      - t_d c_{i}^{\dagger} c_{i+\delta_3}^{}+ h.c. \nonumber\\
     & \quad \quad + i \lambda_{\text soc} \sum_{\langle\!\langle ij \rangle\!\rangle} \sigma v_{ij} c_{i}^{\dagger} c_{j},
\end{align}
where $t_d$ is the nearest-neighbor hopping parameter along the vertical direction ($\delta_3$ in Fig.\ref{fig:honeycomb}(a)). For conciseness, we write the Hamiltonian with a general form,
\begin{align}\label{eq:htbr}
     H^{\sigma} =& 
     \sum_{i\in A} -t_1 (c_{i}^{\dagger} c_{i+\delta_1}^{} + c_{i}^{\dagger} c_{i+\delta_2}^{})
      - t_d c_{i}^{\dagger} c_{i+\delta_3}^{}+ h.c. \nonumber\\
     &\quad + i \lambda_{\text soc} \sum_{\langle\!\langle ij \rangle\!\rangle} \sigma v_{ij} c_{i}^{\dagger} c_{j} - t_3 \sum_{\langle\!\langle\!\langle ij \rangle\!\rangle\!\rangle}c_{i}^{\dagger} c_{j}^{}.
\end{align}
In this form, we have the GKM model when $t_d = t_1$ and we have the DKM when $t_3=0.0, t_d \neq t_1$.
Fourier transforming the Hamiltonian in Eq.\eqref{eq:htbr} to momentum space, we obtain
$H^{\sigma} = \sum_{\bfk} \psi_{\bfk}^\dagger {H}_{\bfk} \psi_{\bfk}^{}$ with 
$\psi_{\bfk}^{}=(c_{{\bf k}A}, c_{{\bf k}B})^T$, where $c_{\mathbf{k}A}$ and $c_{\mathbf{k}B}$ define annihilation operators on the two basis sites in the unit cell shown in Fig.\ref{fig:honeycomb}(a). In the following, we focus on the spin-$\uparrow$ Hamiltonian only,
\begin{align}\label{eq:htbk}
    {H}_{\bfk\uparrow} =&
        \begin{pmatrix}
0 & - f_1({\bf k}) - t_3 f_3({\bf k}) \\
-f_1^*({\bf k}) - t_3 f_3^*({\bf k}) & 0
        \end{pmatrix}
  \nonumber\\
+& \begin{pmatrix}
-2\lambda_{so} g({\bf k}) & 0 \\
0 & 2\lambda_{so} g({\bf k}) 
\end{pmatrix}\;,
\end{align}
where $g(\bfk) \equiv -\sin (\bfk\cdot \bfa_1) + \sin(\bfk\cdot\bfa_2) + \sin(\bfk \cdot {\bfa}_1 - \bfk \cdot {\bfa}_2)$, $f_1(\bfk) = t_d + t_1 e^{-i \bfk \cdot {\bfa}_1} + t_1 e^{-i {\bf k} \cdot {\bfa}_2}$, and $f_3({\bf k}) = e^{-i \bfk\cdot(\bfa_1 + \bfa_2)} + 2 \cos(\bfk\cdot\bfa_1 - \bfk\cdot\bfa_2)$. 
The translational vectors are $\mathbf{a}_{1}=( \sqrt{3}/2,3/2)a$, $\mathbf{a}_{2}=(-\sqrt{3}/2,3/2)a$, and
$\mathbf{a}_{3}=\mathbf{a}_{1}-\mathbf{a}_{2} = (\sqrt{3},0)a$, with $a$ the lattice constant.
The reciprocal-lattice primitive vectors can be chosen as $\bfb_1 = (\sqrt{3}, 1)2\pi/(3a)$, and $\bfb_2 = (-\sqrt{3}, 1)2\pi/(3a)$.
For the GKM model, the gap opened at the ${\bf \Gamma}$ point is $|6(t_1 + t_3)|$, the $\bfK, \bfK'$ points are $|6\sqrt{3} t_2|$, and the $\bfM_{1,2,3}$ points are $2|t_1-3t_3|$.
In this paper, we fix $t_1 = 1.0, t_2 = -0.3$ to make sure the equilibrium system band gap is situated at the $M$ points.
\begin{figure}[t]
\includegraphics[width=0.495\linewidth]{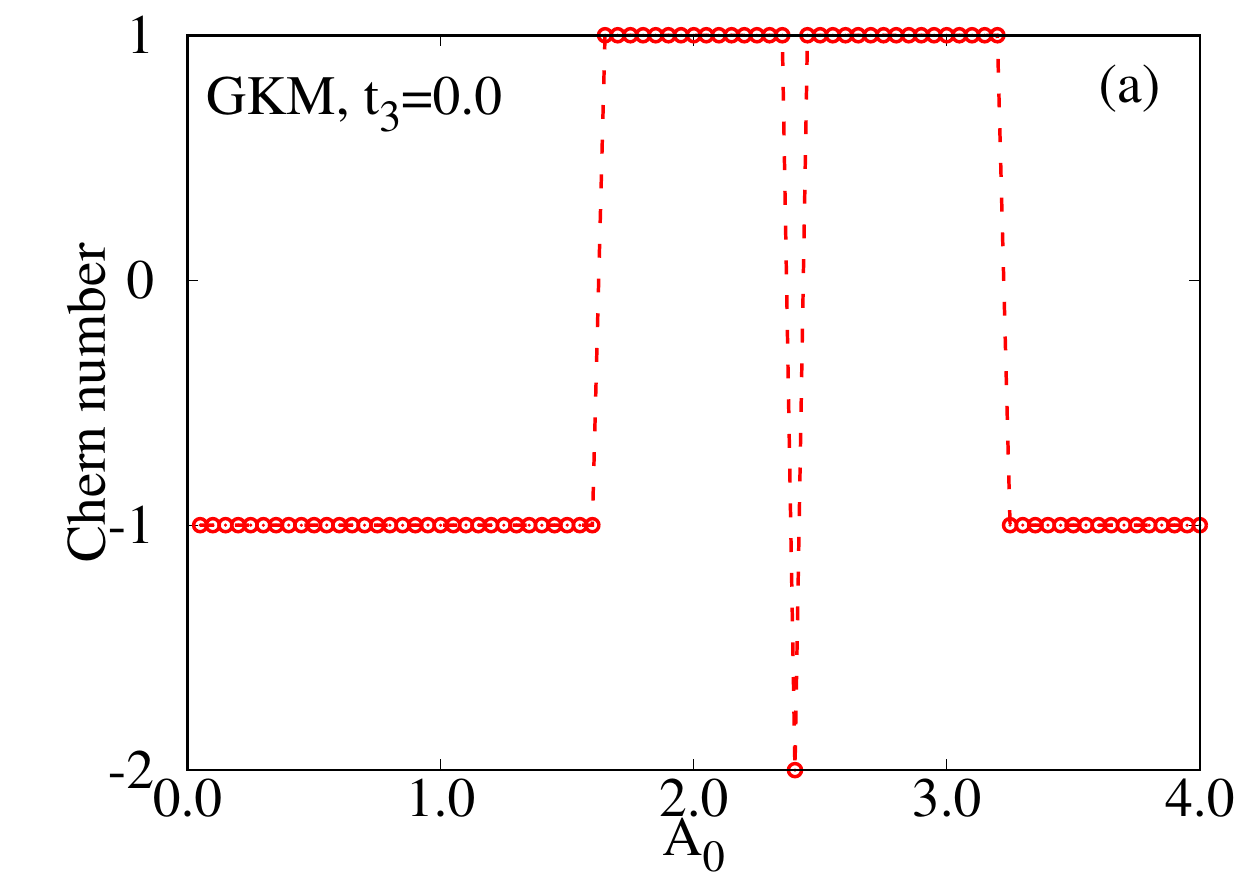}  %
\includegraphics[width=0.495\linewidth]{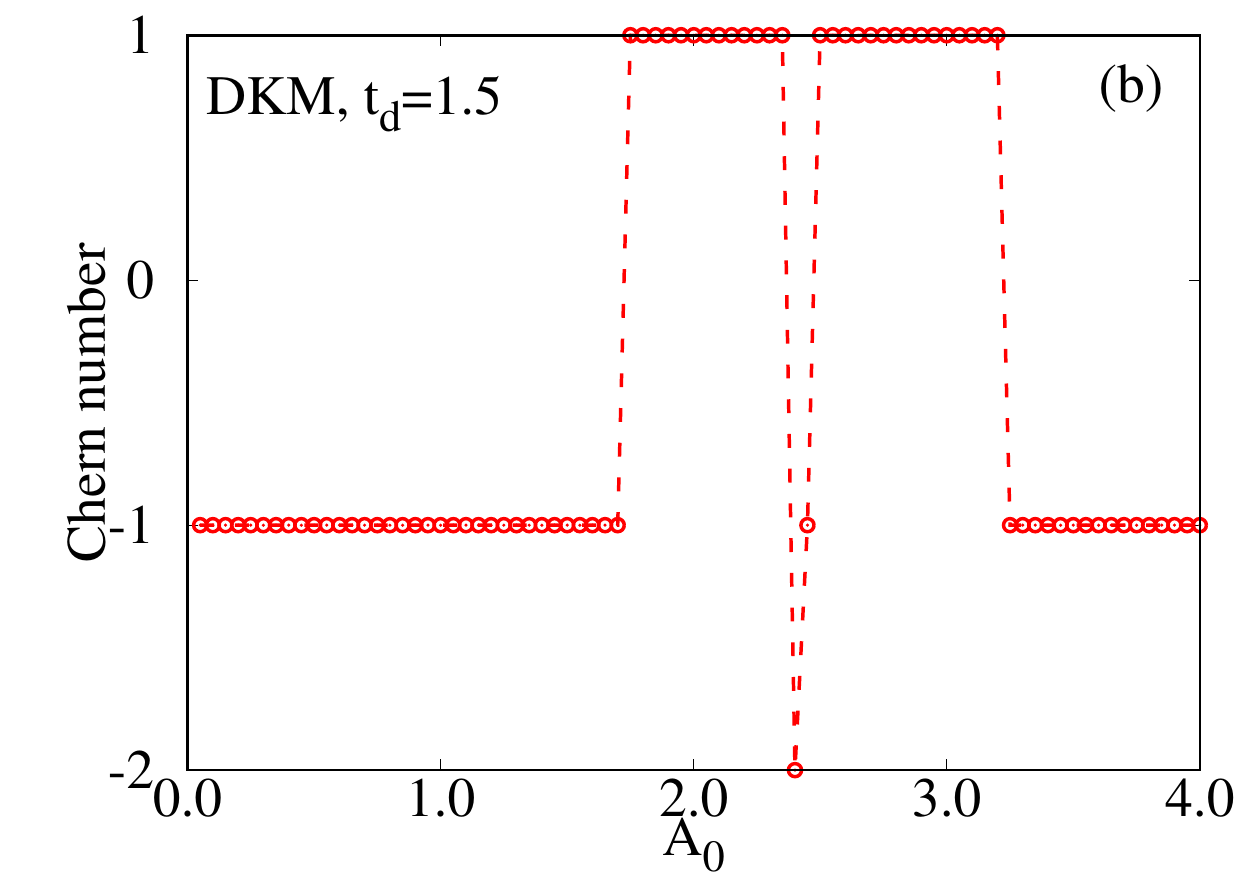}  %
\includegraphics[width=0.495\linewidth]{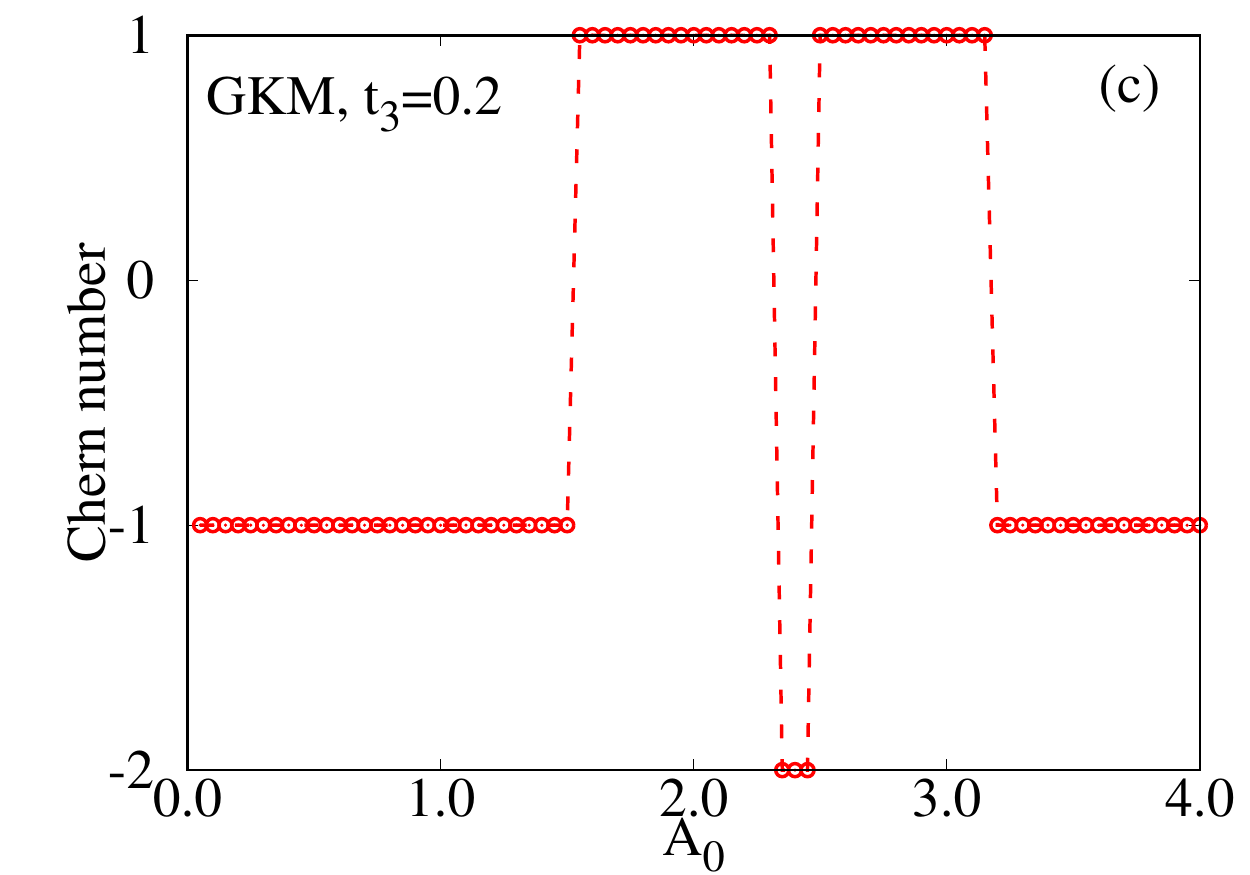}  %
\includegraphics[width=0.495\linewidth]{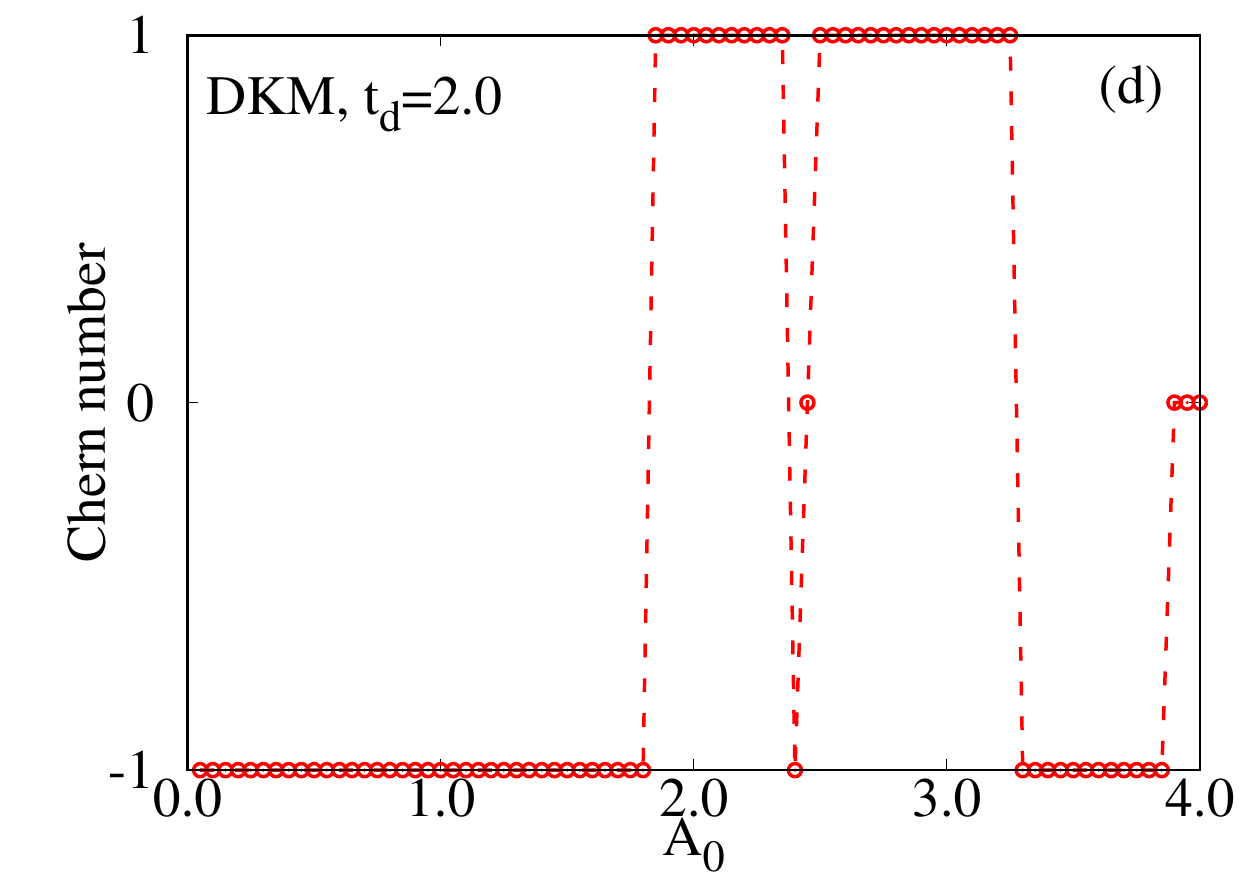}  %
\includegraphics[width=0.495\linewidth]{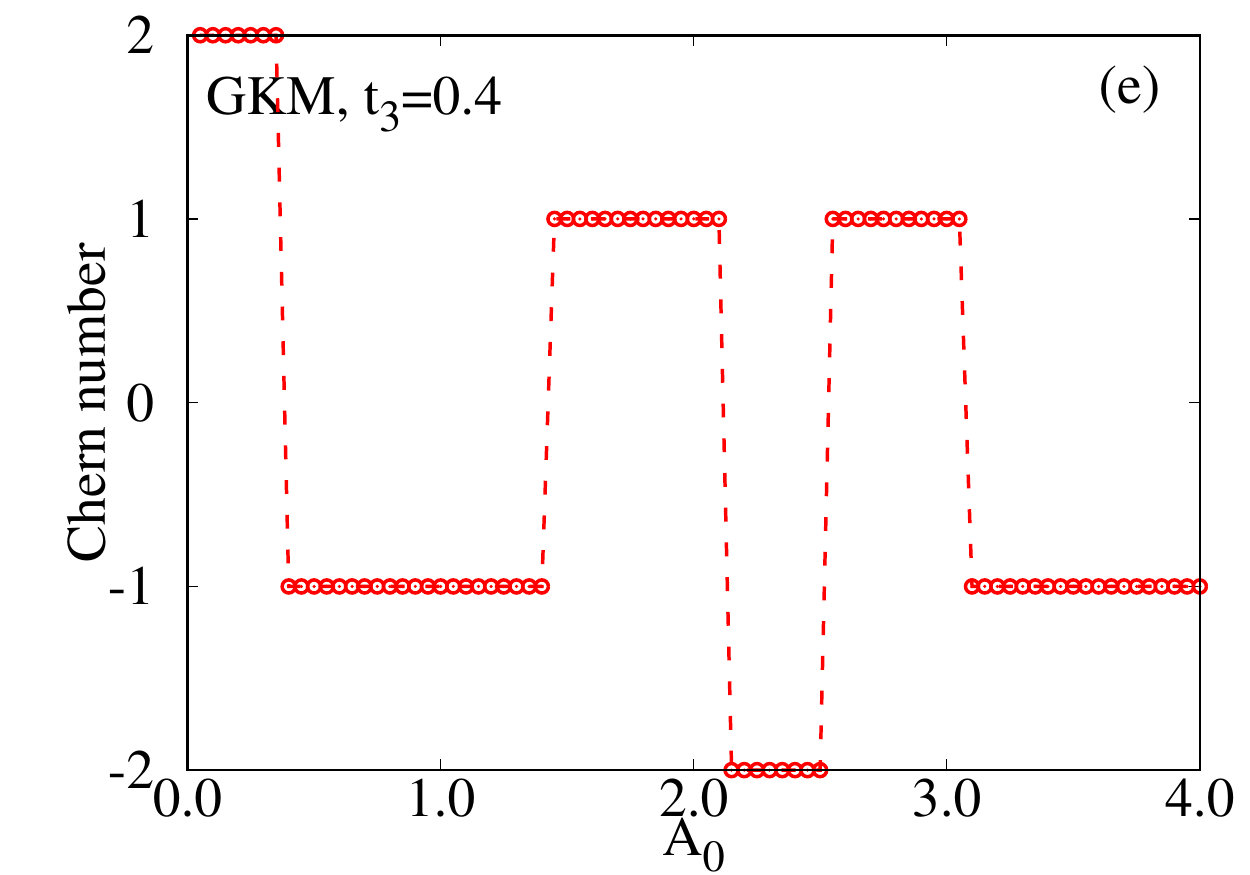}  %
\includegraphics[width=0.495\linewidth]{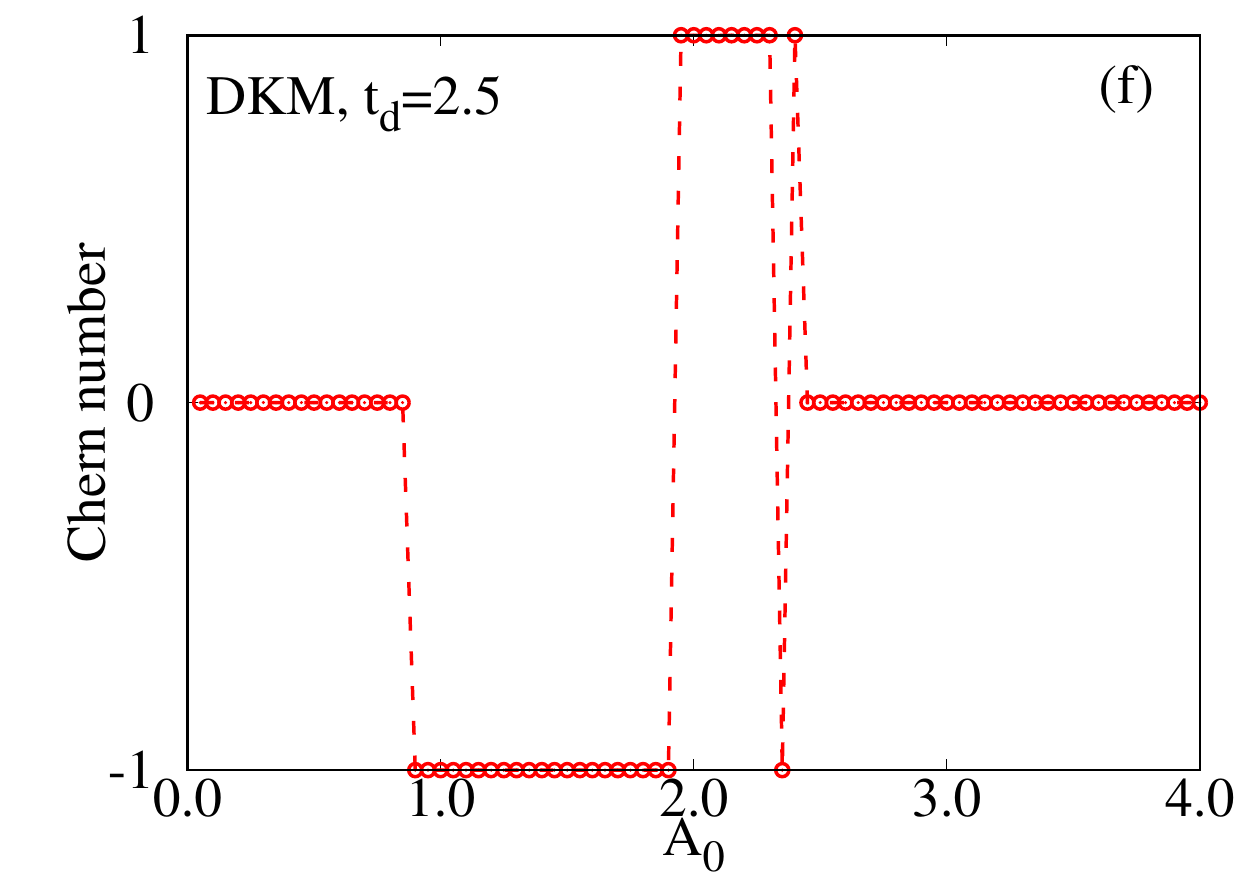}  %
\caption{(Color online) The spin Chern number as a function of laser amplitude $A_0$ for the generalized Kane-Mele model [Eq. \eqref{eq:htbrGKM}] with $t_3=0.0$ (a), $t_3=0.20$ (c), $t_3=0.40$ (e) and the dimerized Kane-Mele model [Eq. \eqref{eq:htbrDKM}] with $t_d=1.5$ (b), $t_d=2.0$ (d), $t_d=2.50$ (f). The remaining parameters are nearest-neighbor hopping $t_1=1.0$, spin-orbit coupling $\lambda=0.3$, and laser frequency $\Omega=10.0$. All the calculations are done with 2500 $\textbf{k}$-points in the first Brillouin zone and 9 Floquet copies.
}
\label{fig:spin-chern}
\end{figure}

\begin{figure*}[t]
\includegraphics[width=0.99\linewidth]{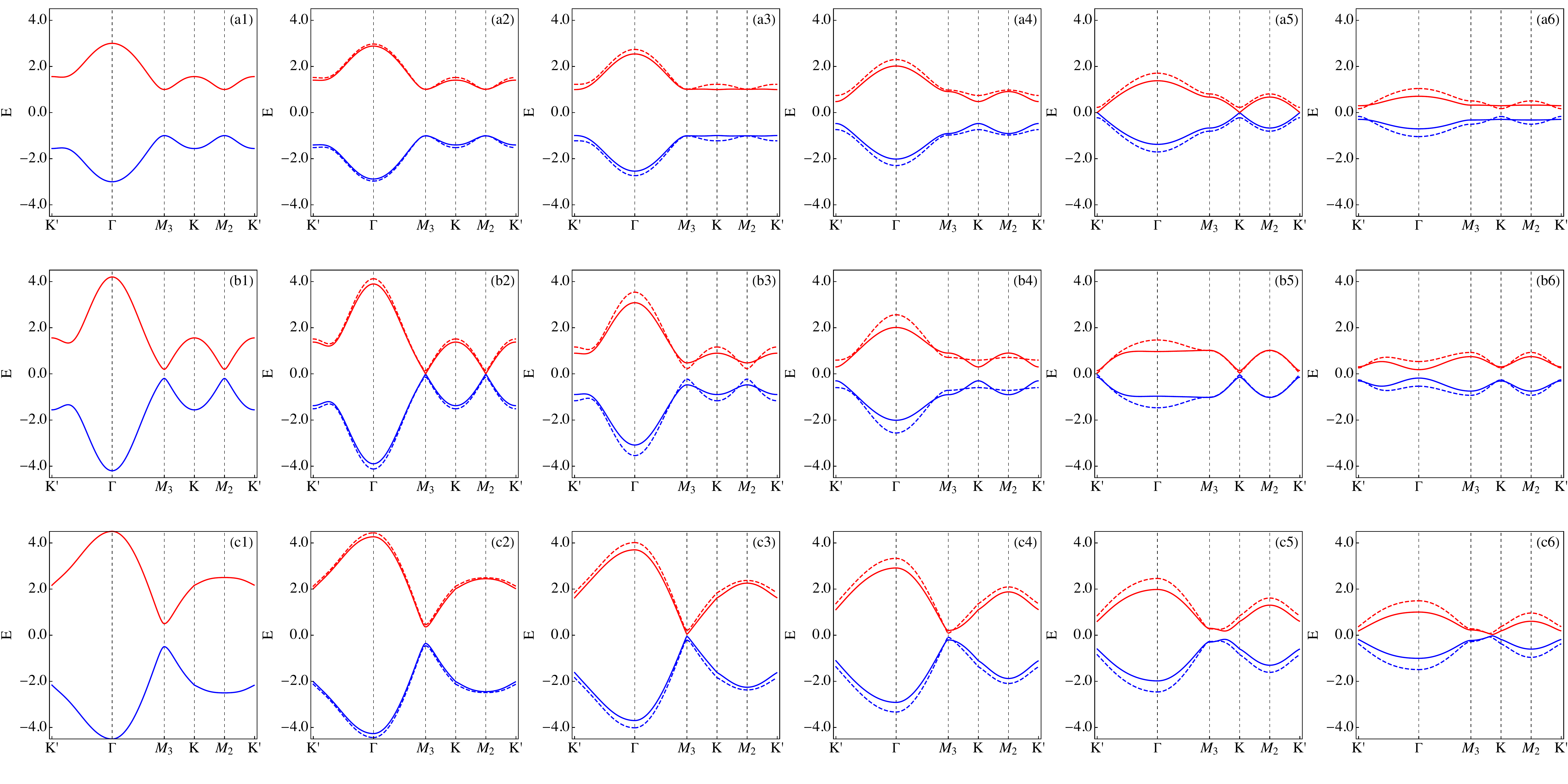}  %
\caption{(Color online) The Floquet-Bloch band structure of the Kane-Mele model [Eq. \eqref{eq:htbrGKM} with $t_3=0.0$] with parameters $t_1=1.0, \lambda=0.3$ and $\Omega=10.0$, 
(a1) $A_0 = 0.0$, (a2) $A_0 = 0.4$, (a3) $A_0 = 0.8$, (a4) $A_0 = 1.2$, (a5) $A_0 = 1.6$, (a6) $A_0 = 2.0$. 
The Floquet-Bloch band structure of the generalized Kane-Mele model [Eq. \eqref{eq:htbrGKM}] with parameters $t_1=1.0$, $\lambda=0.3$, $t_3 = 0.4$ and $\Omega=10.0$ are shown in middle panes, (b1) $A_0 = 0.0$, (b2) $A_0 = 0.4$, (b3) $A_0 = 0.8$, (b4) $A_0 = 1.2$, (b5) $A_0 = 1.6$, (b6) $A_0 = 2.0$.
The Floquet-Bloch band structure of the dimerized Kane-Mele model [Eq. \eqref{eq:htbrDKM}] with parameters $t_1=1.0$, $\lambda=0.3$, $t_d = 2.4$ and $\Omega=10.0$ are shown in bottom panes, (c1) $A_0 = 0.0$, (c2) $A_0 = 0.4$, (c3) $A_0 = 0.8$, (c4) $A_0 = 1.2$, (c5) $A_0 = 1.6$, (c6) $A_0 = 2.0$.  Note the solid lines denote the Floquet band structure with laser intensity $A_0$ shown above; the dashed lines show the data for $A_0 - 0.2$ for comparison.
}
\label{fig:Floquetband}
\end{figure*}

When Eq.\eqref{eq:htbr} is exposed to a normally incident laser field, the time-dependent Hamiltonian can be expressed as,
\begin{align}\label{eq:htbr-t}
     H(t) =& 
       \sum_{i\in A} \left[-t_1 c_{i}^{\dagger} c_{i+\delta_1}^{} - t_1 c_{i}^{\dagger} c_{i+\delta_2}^{}
      - t_d c_{i}^{\dagger} c_{i+\delta_3}^{}\right]e^{-i A_{ij}(t)}  \nonumber\\
      &- t_3 \sum_{\langle\!\langle\!\langle ij \rangle\!\rangle\!\rangle} c_{i}^{\dagger} c_{j}^{}e^{-i A_{ij}(t)} + h.c. \nonumber\\
     &+ i \lambda_{\text soc} \sum_{\langle\!\langle ij \rangle\!\rangle} \sigma v_{ij} c_{i}^{\dagger} c_{j}e^{-i A_{ij}(t)}\;,  
\end{align}
where $A_{ij}(t) = {\bf A}(t)\cdot({\bf R}_j - {\bf R}_i) $, ${\bf A}(t)=A_0[\sin(\Omega t), \cos(\Omega t)]$ is the vector potential with $A_0$ the amplitude and $\Omega$ the frequency of the laser.
The relation $\bfR_j = \bfR_i + \delta_i$ with $i=1,2,3$ for each term holds.
In Eq.\eqref{eq:htbr-t}, we set Planck's constant $\hbar=1$, the speed of light $c=1$, the charge of the electron $e=1$, and adopt the Coulomb gauge by setting the scalar potential $\phi=0$. We ignore the tiny effect of the magnetic field of the laser field.  The units of energy are expressed in terms of the nearest-neighbor hopping amplitude $t_1$, for $t_1=1$,
\begin{align}\label{eq:htbk-t}
         {H}_{\mathbf{k}\uparrow}(t) =&
        \begin{pmatrix}
0 & - f_1(\bfk,t)  \\
- f_1^*(\bfk,t) & 0
        \end{pmatrix}
  \nonumber\\
+& \begin{pmatrix}
-i\lambda_{so} g({\bfk,t}) & - t_3 f_3(\bfk,t) \\
 - t_3 f_3^*(\bfk,t) & i\lambda_{so} g({\bfk,t}) 
\end{pmatrix},
\end{align}
where 
\begin{align}
g({\bfk,t}) &\equiv e^{i\bfk\cdot\bfa_1-i\bfA(t)\cdot\bfa_1}-e^{-i\bfk\cdot\bfa_1+i\bfA(t)\cdot\bfa_1} \nonumber\\
&-e^{i\bfk\cdot\bfa_2-i\bfA(t)\cdot\bfa_2}+e^{-i\bfk\cdot\bfa_2+i\bfA(t)\cdot\bfa_2} \nonumber\\
&-e^{i\bfk\cdot\bfa_3-i\bfA(t)\cdot\bfa_3}+e^{-i\bfk\cdot\bfa_3+i\bfA(t)\cdot\bfa_3},
\end{align}
\begin{align}
f_1({\bfk},t) = t_d e^{-i\bfA(t)\cdot\delta_3} 
&+ t_1 e^{-i \bfk\cdot\bfa_1}e^{-i\bfA(t)\cdot\delta_1} \nonumber\\
&+ t_1 e^{-i \bfk\cdot\bfa_2}e^{-i\bfA(t)\cdot\delta_2},
\end{align}
and
\begin{align}
f_3({\bfk},t) &= e^{-i {\bf k}\cdot({\bfa}_1 + {\bfa}_2)}e^{i\bfA(t)\cdot2\delta_3} 
+ e^{i {\bf k}\cdot\bfa_3} e^{-i\bfA(t)\cdot(\bfa_1+\delta_2)} \nonumber\\ 
&+ e^{-i {\bf k}\cdot\bfa_3} e^{-i\bfA(t)\cdot(\bfa_2+\delta_1)}.
\end{align}
\section{Floquet Theory}
\label{sec:Floquet}
In this paper, we illuminate the system with monochromatic (single frequency) light, which renders the Hamiltonian time-periodic: $H(t) = H(t+T)$ where $T$ is the period of the laser drive. Hence, Floquet's theorem is applicable. The Floquet eigenfunction in real space for the time-periodic Hamiltonian can be expressed as,
\begin{equation}
     |\Psi_{\alpha}(t)\rangle = e^{-i\epsilon_{\alpha} t} |\phi_{\alpha}(t)\rangle,
\end{equation}
where $|\phi_{\alpha}(t)\rangle = |\phi_{\alpha}(t+T)\rangle$ are the Floquet quasi-modes and 
$\epsilon_{\alpha}$ is the corresponding quasi-energy for band $\alpha$.
Substituting the wave function above into the time-dependent Schr\"odinger equation, 
and defining the Floquet Hamiltonian operator as $\mathcal{H}(t) = H(t) - i{\partial}/{\partial t}$, one finds
\begin{equation}
     \mathcal{H}(t) |\phi_{k\alpha}(t)\rangle = \epsilon_{\alpha} |\phi_{\alpha}(t)\rangle.
\end{equation}
Here we restrict the quasienergy to be in the first Floquet zone, {\it i.e.}, $-\Omega/2 < \epsilon_{\alpha} < \Omega/2$. (Note that we have made use of a spin-independent coupling to the laser field so that all bands are 2-fold degenerate.  Henceforth, we suppress the spin degeneracy.)
Solving for the Floquet states in Fourier space,
\begin{equation}
     |\phi_{\alpha}(t)\rangle = \sum_{m} e^{i m \Omega t} |\tilde{\phi}_{\alpha}^m\rangle,
\end{equation}
where $m=0,\pm 1, \pm 2, \cdots $ and $|\tilde{\phi}_{k\alpha}^m\rangle$ is a real space vector which obeys,
\begin{align}
     \sum_{m} (H_{nm} + m\Omega\delta_{nm}) |\tilde{\phi}_{\alpha}^m\rangle = \epsilon_{\alpha} |\tilde{\phi}_{\alpha}^m\rangle,
\end{align}
with matrix elements of the Floquet Hamiltonian written as,
\begin{align}
    H_{nm}= \frac{1}{T}\int_{0}^{T} dt e^{-i(n-m)\Omega t} H(t). 
\end{align}
Here $m$ and $n$ are integers ranging from $-\infty$ to $\infty$. Thus, the Floquet matrix is an infinite-dimensional time-independent matrix.  In this paper, we consider the laser frequency to be comparable to or larger than the bandwidth of the system, so a
truncation of the components to be in $m,n=-2, -1,0,1,2$ is a good approximation.  We have numerically verified that including a larger range of $m,n$ has a very small numerical impact on our results.

For circularly polarized light with vector potential $\bfA(t)=A_0[\sin(\Omega t), \cos(\Omega t)]$, the matrix elements of the Floquet-Bloch Hamiltonian are
\begin{align}
     H^{ij}_{nm} = \frac{1}{T} \int_0^T dt e^{-i(n-m)\Omega t}  \exp[-i A_{ij}(t)] H^{ij},
\end{align}
from the expression with the general form,
\begin{align}
      f_{nm} = \frac{1}{T} \int_0^T dt e^{-i(n-m)\Omega t}  \exp[-i \bfA(t)\cdot{\bf d}].
\end{align}
Here we used ${\bf d}={\bf R}_j-{\bf R}_i$, and define $d^x /|{\bf d}| = \cos\theta$, $d^y/|{\bf d}|= \sin\theta$. For nearest-neighbor hopping terms, $|{\bf d}|=1$, $\theta=\pm 5\pi/6, \pm \pi/6, \mp \pi/2$.
Substituting the vector potential into the above equation gives,
\begin{align}
          &\frac{1}{T} \int_0^T dt e^{-i(n-m)\Omega t}  \exp[-i A_0 (d^x \sin\Omega t + d^y \cos\Omega t)] \nonumber\\
      =&\mathcal{J}_{m-n}(A_0 |{\bf d}|) \exp[i(n-m)\theta],
\end{align}
where $\mathcal{J}_{n}(x)$ is the Bessel function of first kind.
\section{Spin Chern number for the disorder-free system}
\label{sec:clean}
\subsection{Spin Chern number and Floquet band structure}
In Fig.\ref{fig:spin-chern}, we plot the spin Chern number as a function of laser intensity for different third-neighbor hopping parameters $t_3=0.0, 0.2, 0.4$ in the generalized Kane-Mele model [Eq.\eqref{eq:htbrGKM}] and different dimerized hopping parameters $t_d=1.5, 2.0, 2.5$ in the dimerized Kane-Mele model [Eq.\eqref{eq:htbrDKM}]. 

In the equilibrium case (absent the laser, i.e. $A_0=0$) of the GKM model, the system gap is determined by the bands at the $M_{1,2,3}$ points. By tuning the third-neighbor hopping parameter, the transition from topologically non-trivial ($\mathcal{C} = -1$) to topologically trivial ($\mathcal{C} =  2$)  occurs at the critical value of $t_3=1/3$, where the gap at $M_{1,2,3}$ ($C_3$ rotational symmetry is conserved) closes and reopens, inducing a $\pm 3$ change of spin Chern number. 
This is the starting point of the non-equilibrium study shown in Fig.\ref{fig:spin-chern}(a),(c),(e). 

In the DKM model, by comparison, the system gap is determined by the bands at the $M_3$ point. By tuning the dimerized nearest-neighbor hopping, the transition from topologically non-trivial ($\mathcal{C} = -1$) to topologically trivial ($\mathcal{C} =  0$) occurs. 
Increasing the dimerized hopping parameter will close the gap at the $M_3$ point ($C_3$ rotational symmetry is broken), and reopen the gap at the critical value $t_d=2.0$, inducing a change of Chern number $\Delta\mathcal{C} = \pm 1$. 
This is the starting point of the non-equilibrium study in Fig.\ref{fig:spin-chern} (b),(d),(f). 

The spin Chern number shows complicated structure for both the GKM and DKM models when illuminated with a laser. 
Since Fig.\ref{fig:spin-chern}(a),(b),(c),(d) have very similar structure, our analysis of the topological transition will be focused on Fig.\ref{fig:spin-chern}(a),(e),(f). 
The transition at weak laser intensity can be easily understood.
In Fig.\ref{fig:spin-chern}(a), increasing the laser intensity will induce the transition from topologically non-trivial states to topologically trivial states. This transition can be understood by plotting the band structure as a function of laser intensity $A_0$, as in Fig.\ref{fig:Floquetband}(a1-a6). At $A_0=0.0$ (laser absent), the system gap is determined by the energy difference at the $M_{1,2,3}$ points (Fig.\ref{fig:Floquetband}(a1)). Increasing the laser intensity tends to form a flat band in the region $M_3-K-M_2-K'$ (Fig.\ref{fig:Floquetband}(a3)). Further increasing laser intensity will set the $K(K')$ point to determine the band gap (Fig.\ref{fig:Floquetband}(a4)). Increasing the laser intensity still further will close, and then reopen, the gap at the $K$ point (Fig.\ref{fig:Floquetband}(a5-a6)), inducing the Chern number change $\Delta\mathcal{C} = \pm 1$ for $K(K')$.  This explains the topological transition at $A_0=1.5$. 
In Fig.\ref{fig:spin-chern}(e), the first transition at $A_0=0.4$ is induced by the three $M_{}$ points' band closing and reopening (Chern number change $\pm 1$ for each $M$ point) driven by laser coupling, while the second transition at $A_0=1.5$ is because of the $K(K')$ points closing and reopening (Chern number change $\pm 1$ for $K$ or $K'$ point). In Fig.\ref{fig:spin-chern}(f), the first transition at $A_0=0.8$ is due to the $M_3$ point closing and reopening (Chern number change $\pm 1$), and the transition at $A_0=1.8$ is due to the $K(K')$ points closing and reopening (Chern number change $\pm 1$) for $K$ or $K'$ point. The picture can be confirmed by plotting the Floquet band structure with different laser intensities. 

\subsection{Low energy Hamiltonian in the high frequency limit}
\begin{table*}[t] \renewcommand{\arraystretch}{2.5}
  \begin{center}
    \caption{Energy gap at high symmetry point in the theoretical infinite frequency limit.}
    \label{tab:table1}
    \begin{tabular}{|c|c|c|c|c|}
    \hline
          & $\bfK (\bfK')$ & $\bfM_1 (\bfM_2)$ & $\bfM_3$ & $\Gamma$\\
    \hline
      $\GKM$ & $6\sqrt{3}\lambda \mathJ_0(\sqrt{3}A_0)$ & $2\left|t_1 \mathJ_0(A_0) - 3t_3 \mathJ_0(2A_0)\right|$  &  $2\left|t_1 \mathJ_0(A_0) - 3t_3 \mathJ_0(2A_0)\right|$ & $6\left|t_1 \mathJ_0(A_0) + t_3 \mathJ_0(2A_0)\right|$\\
      \hline
      $\DKM$ & $2\sqrt{27\lambda^2 \mathJ_0(\sqrt{3}A_0)^2 + (t_1 - t_d)^2 \mathJ_0(A_0)^2}$ & $2\left|t_d \mathJ_0(A_0)\right|$ & $2\left|(2t_1 - t_d) \mathJ_0(A_0)\right|$ & $2\left|(2t_1 + t_d) \mathJ_0(A_0)\right|$ \\
      \hline
    \end{tabular}
  \end{center}
\end{table*}
The position of the $\bf K$ point is ($4\pi/3\sqrt{3}, 0$) (and symmetry related points), and the low-energy Hamiltonian at the high-frequency limit is given by, 
\begin{align}
     H_{0} =  \begin{pmatrix}
     3\sqrt{3}\lambda\mathJ_0(\sqrt{3}A_0) & (t_1 - t_d) \mathJ_0(A_0) \\
     (t_1 - t_d) \mathJ_0(A_0) & -3\sqrt{3}\lambda\mathJ_0(\sqrt{3}A_0)
     \end{pmatrix}.
\end{align}
For the generalized Kane-Mele model, we have $t_1 = t_d$. Then
the eigenvalues will be 
\begin{align}
     E_{\pm} = \pm 3\sqrt{3}\lambda \mathJ_0(\sqrt{3}A_0),
\end{align}
which depend on only the spin-orbit coupling $\lambda$ and scaled by Bessel function $\mathcal{J}(\sqrt{3}A_0)$.
For the dimerized Kane-Mele model, the Hamiltonian is independent of the third-neighbor hopping terms $t_3$. The eigenvalues are 
\begin{align}
     E_{\pm} = \pm \sqrt{27\lambda^2 \mathJ_0(\sqrt{3}A_0)^2 + (t_1 - t_d)^2 \mathJ_0(A_0)^2}. 
\end{align}

The position of ${\bfM}_3$ point is ($0, 2\pi/3$), and the low-energy Hamiltonian up to second order in $A_0$ is given by, 
\begin{align}
     H_{0} = \begin{pmatrix}
     0 & f_{\bfM_3}  \\
     f_{\bfM_3} & 0
     \end{pmatrix},
\end{align}
with $f_{\bfM_3} =(2t_1 - t_d) \mathJ_0(A_0) - 3t_3 \mathJ_0(2A_0)$. The eigenvalues are
\begin{align}
     E_{\pm} &= \pm \left|(2t_1 - t_d) \mathJ_0(A_0) - 3t_3 \mathJ_0(2A_0)\right|.
\end{align}
For the generalized Kane-Mele model, we have $t_1 = t_d$, then
the eigenvalues will be 
\begin{align}
    E_{\pm} &= \pm \left|t_1 \mathJ_0(A_0) - 3t_3 \mathJ_0(2A_0)\right|.
\end{align}
For the dimerized Kane-Mele model, the Hamiltonian is independent of third-neighbor hopping terms $t_3$. The eigenvalues are 
\begin{align}
     E_{\pm} &= \pm \left|(2t_1 - t_d) \mathJ_0(A_0)\right|.
\end{align}
\begin{figure}[h]
\includegraphics[width=0.495\linewidth]{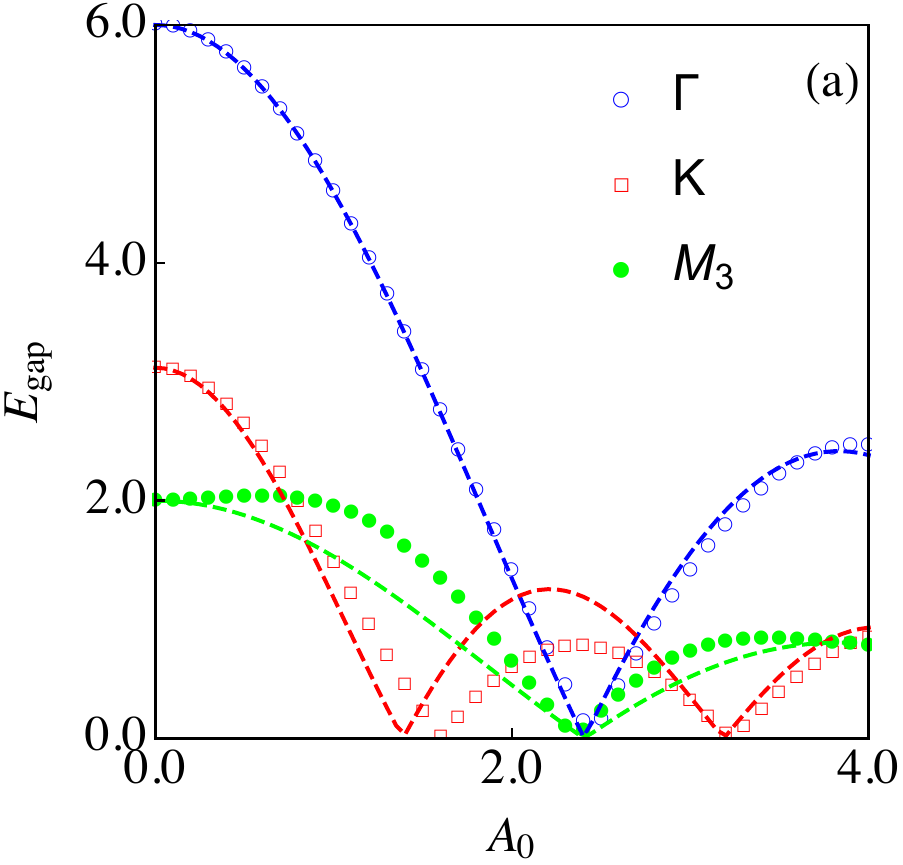}  %
\includegraphics[width=0.495\linewidth]{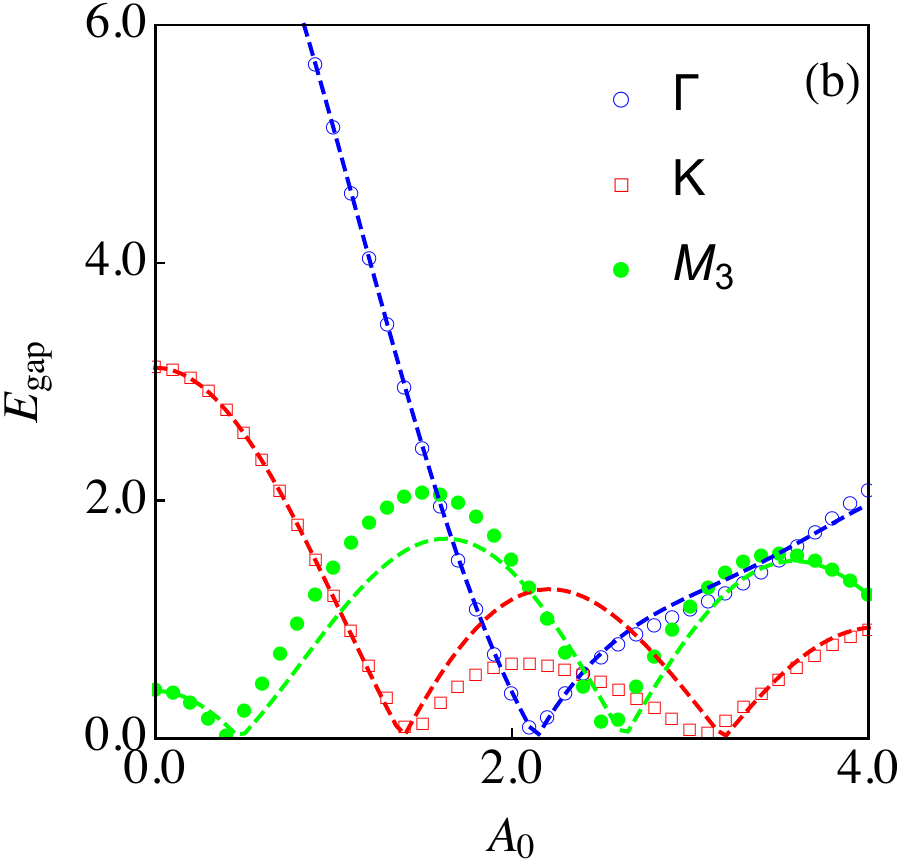}  %
\includegraphics[width=0.495\linewidth]{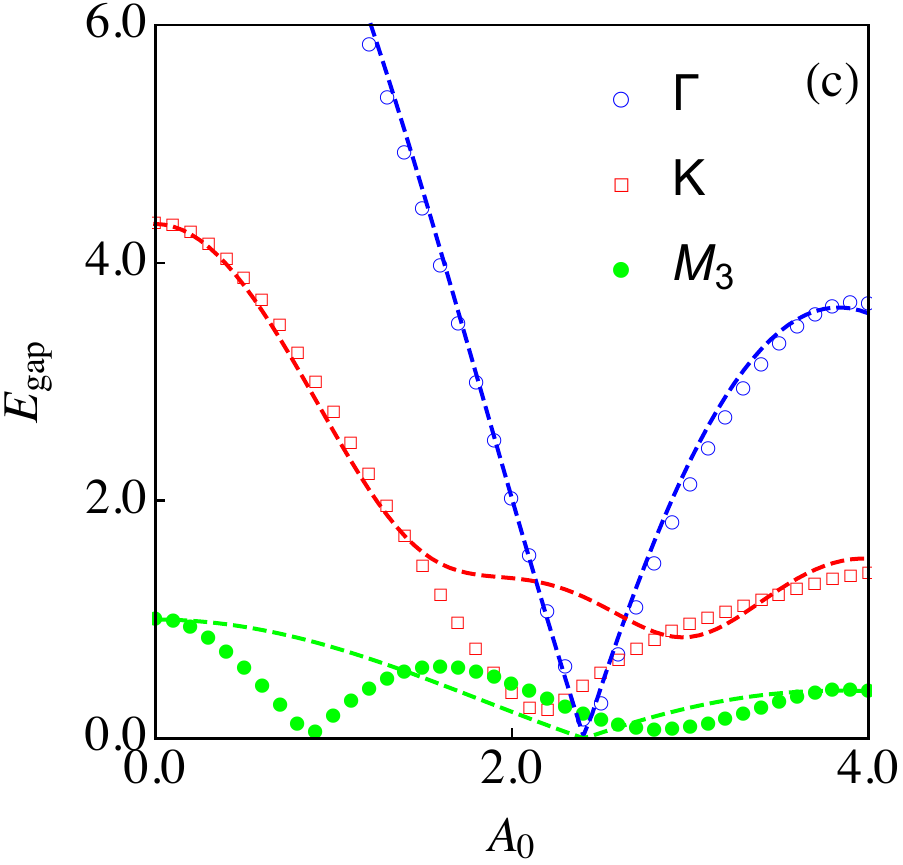}  %
\includegraphics[width=0.495\linewidth]{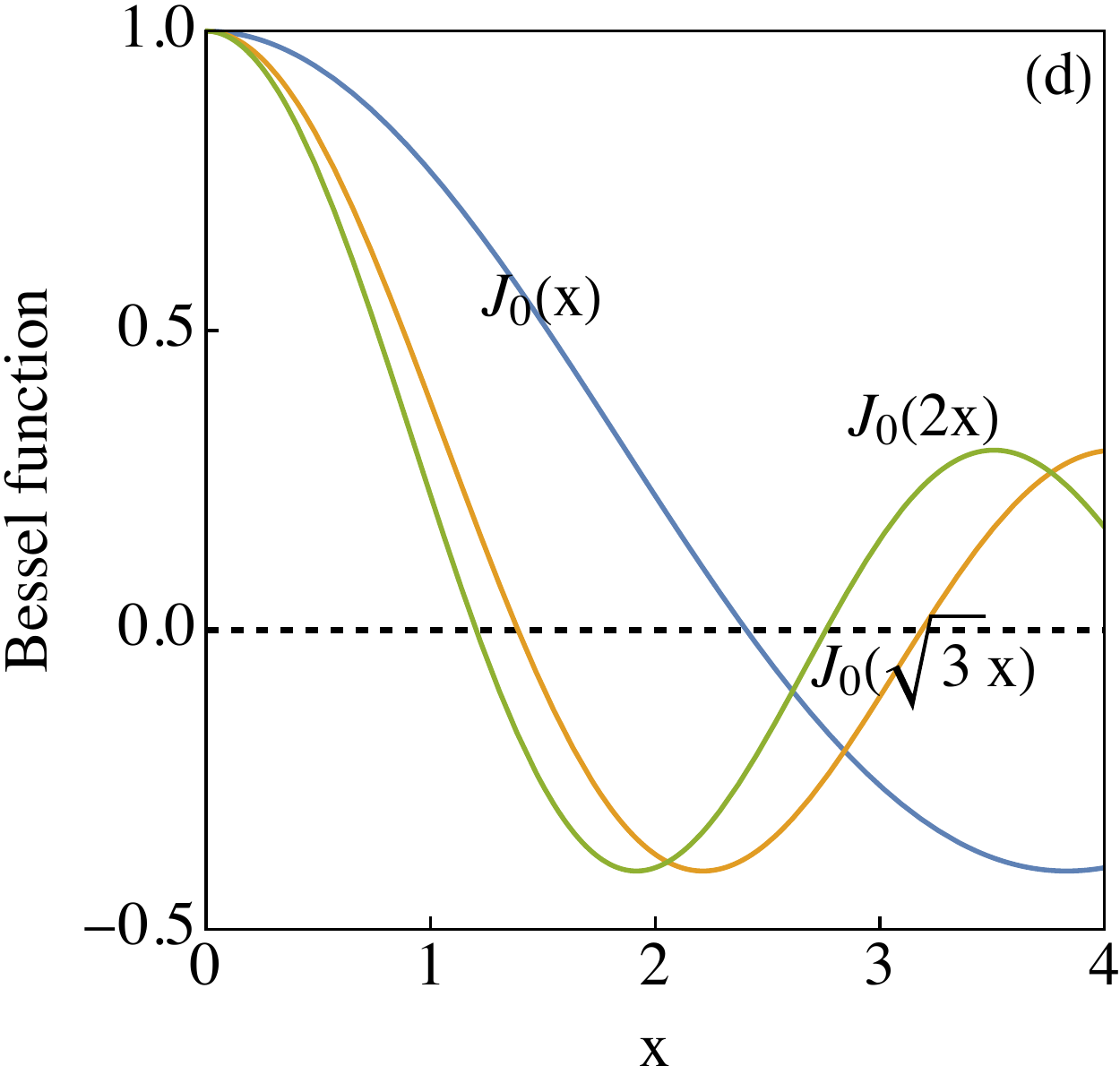}  %
\caption{(Color online) The energy gaps at high symmetry points are plotted with dots. The dashed line indicates the energy gap in the theoretical infinite-frequency limit. (a) Generalized Kane-Mele model with $t_3 = 0.0$. (b) Generalized Kane-Mele model with $t_3 = 0.4$. (c) Dimerized Kane-Mele model with $t_d = 2.5$. (d) The zero-$th$ order Bessel function of first kind used in the infinite frequency limit.
}
\label{fig:gap}
\end{figure}
\begin{figure*}[t]
\includegraphics[width=0.16\linewidth]{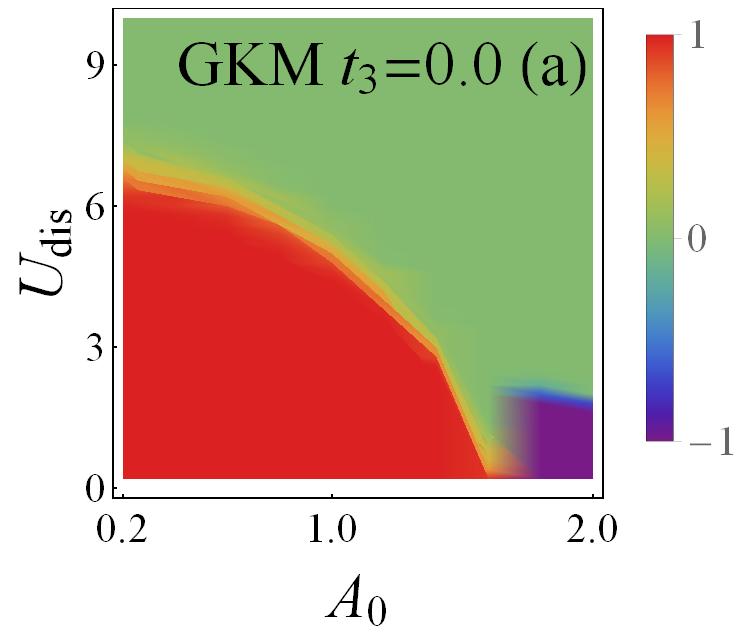}  %
\includegraphics[width=0.16\linewidth]{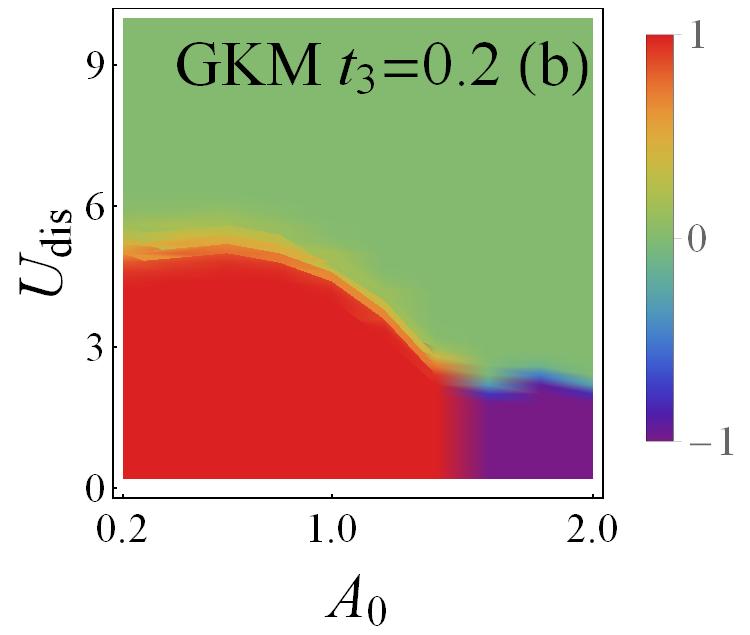}  %
\includegraphics[width=0.16\linewidth]{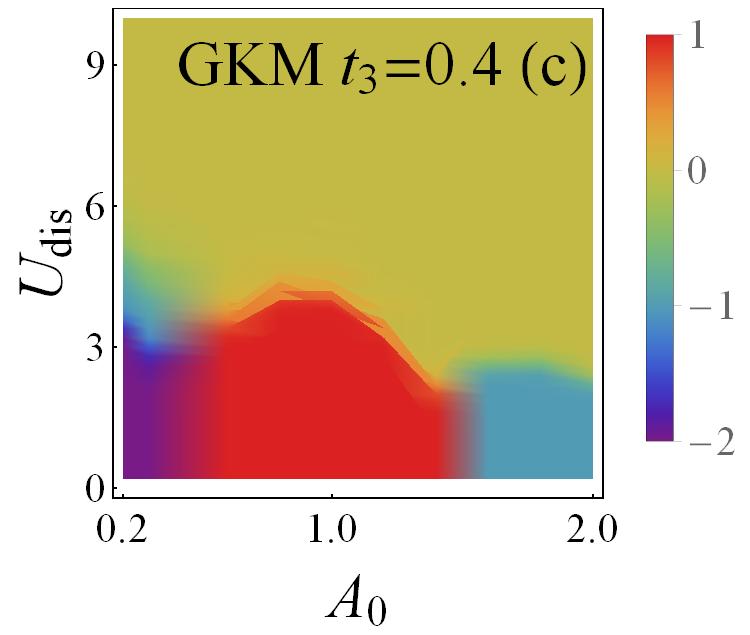}  %
\includegraphics[width=0.16\linewidth]{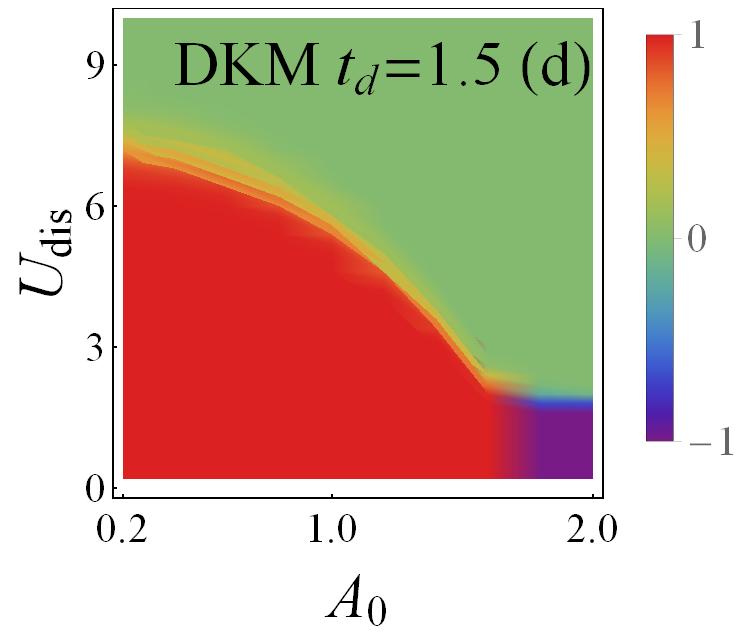}  %
\includegraphics[width=0.16\linewidth]{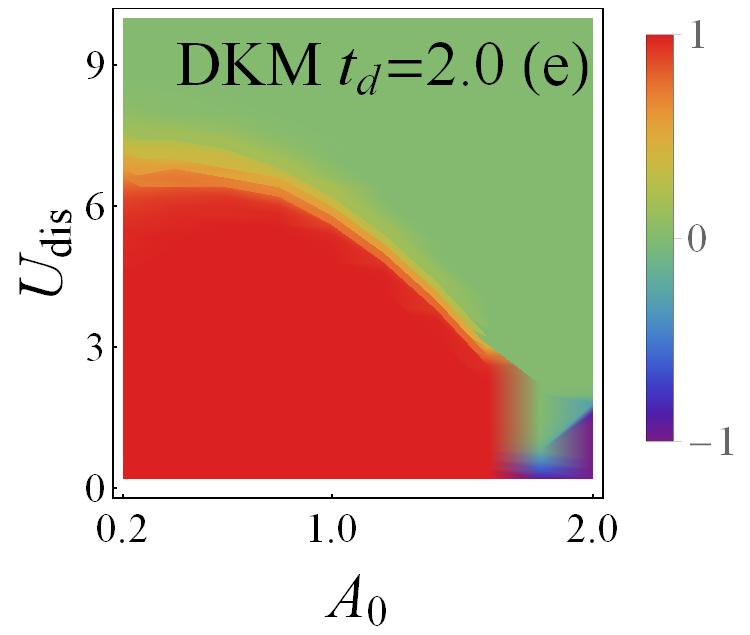}  %
\includegraphics[width=0.16\linewidth]{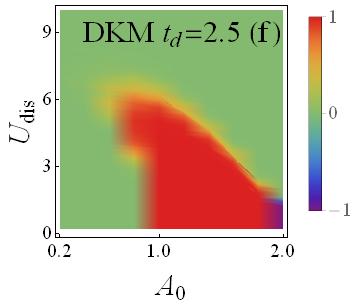}  %
\includegraphics[width=0.33\linewidth]{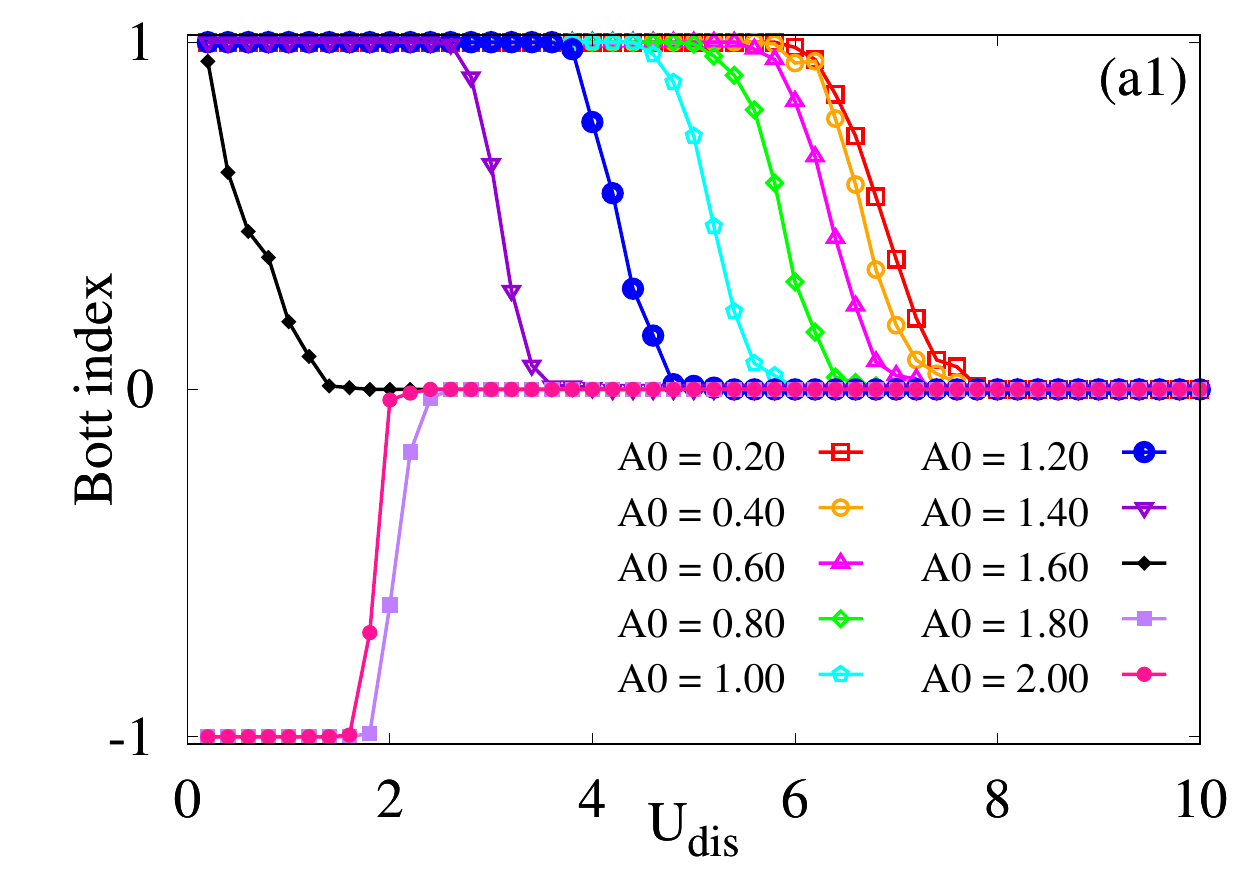}  %
\includegraphics[width=0.33\linewidth]{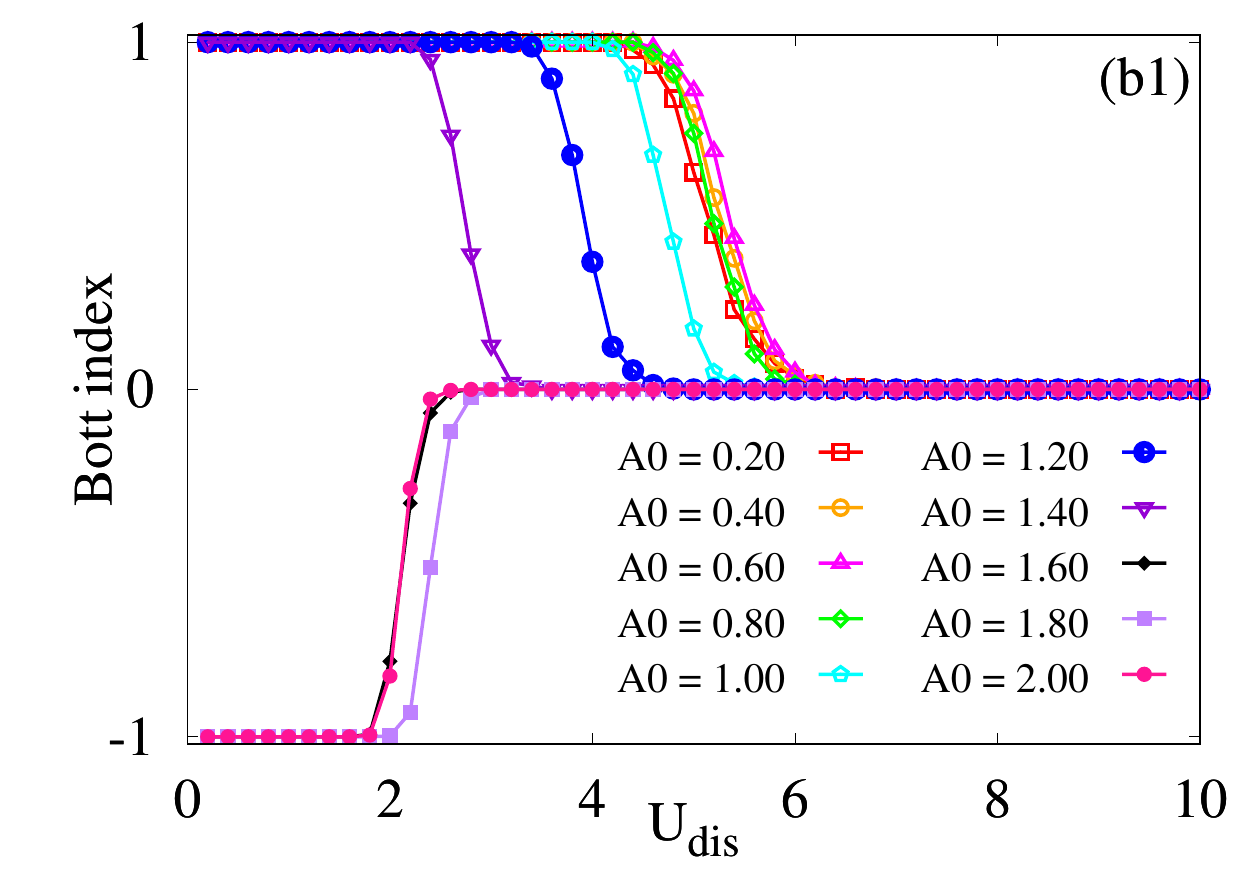}  %
\includegraphics[width=0.33\linewidth]{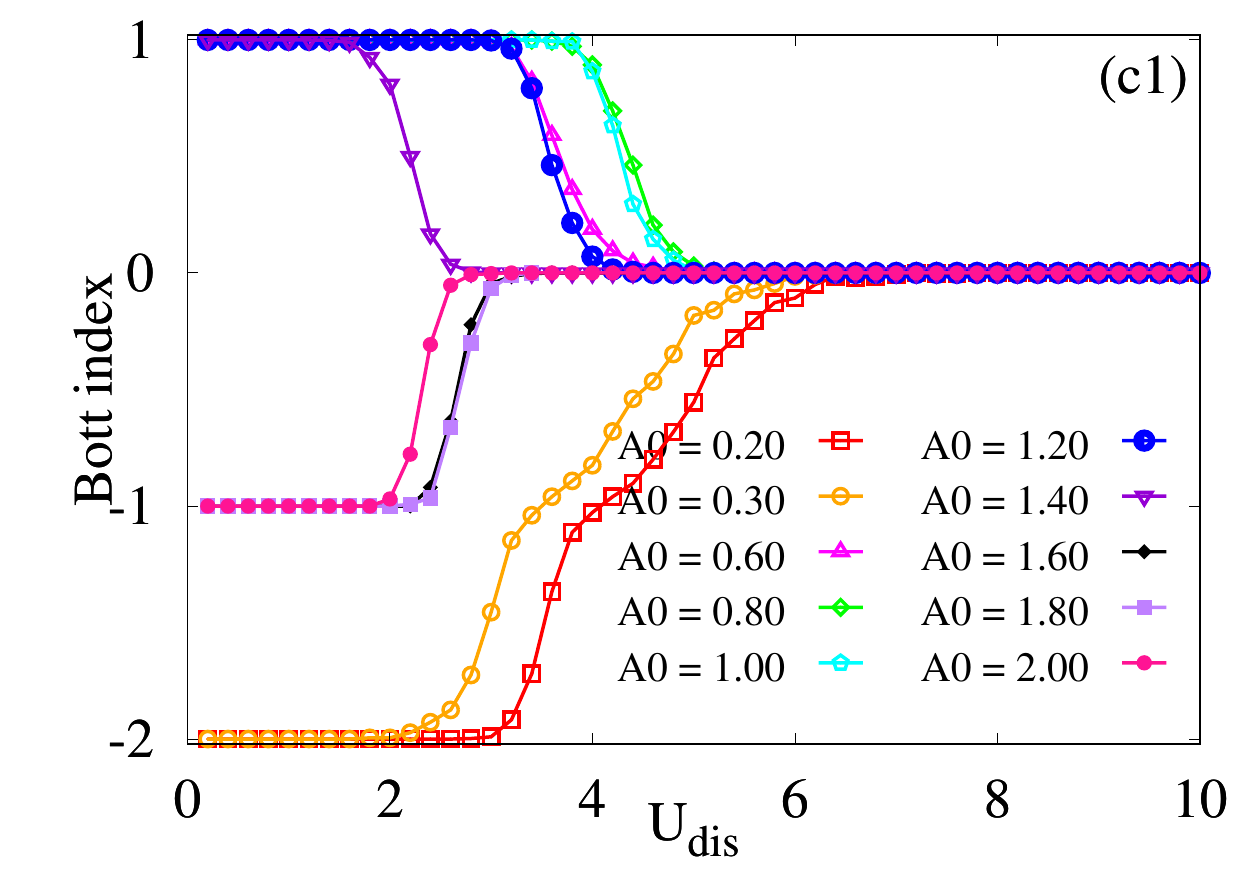}  %
\includegraphics[width=0.33\linewidth]{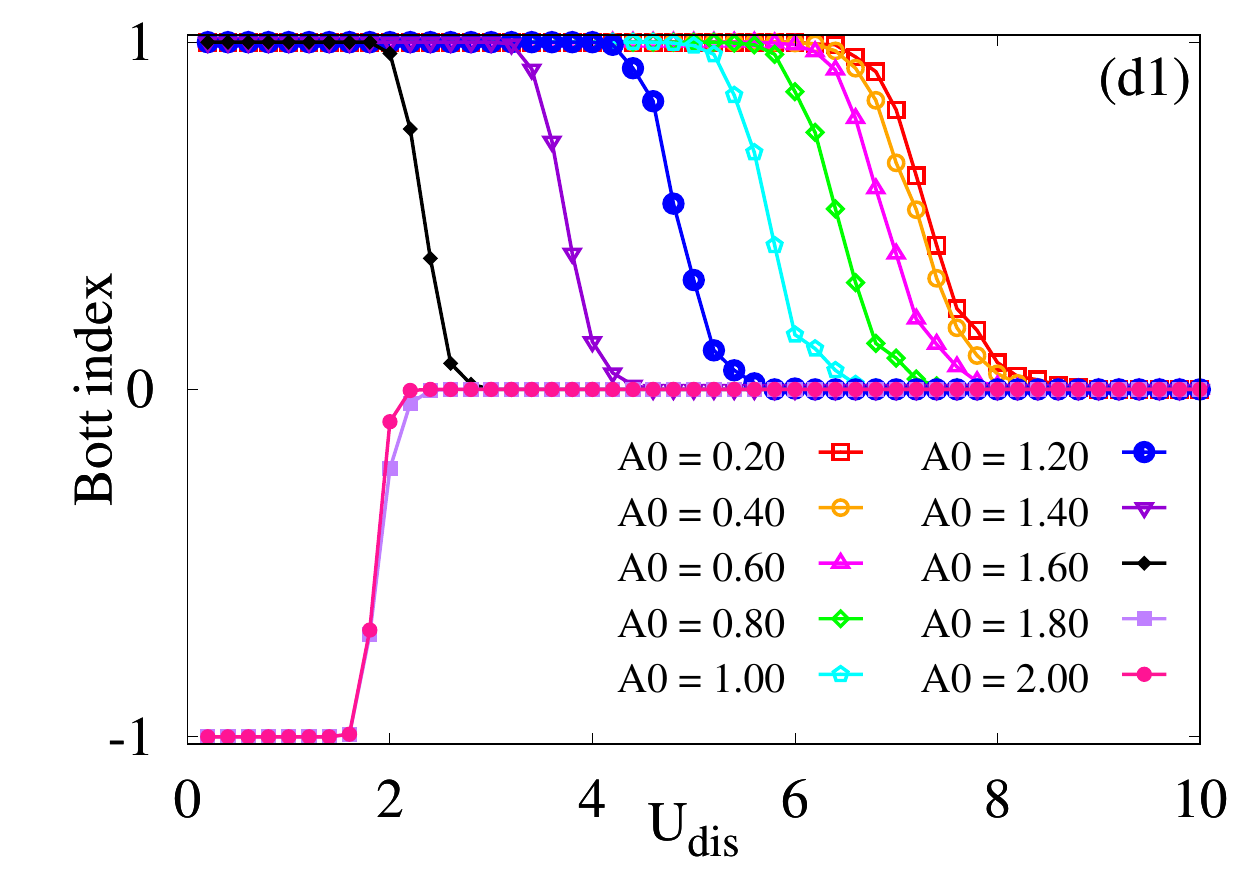}  %
\includegraphics[width=0.33\linewidth]{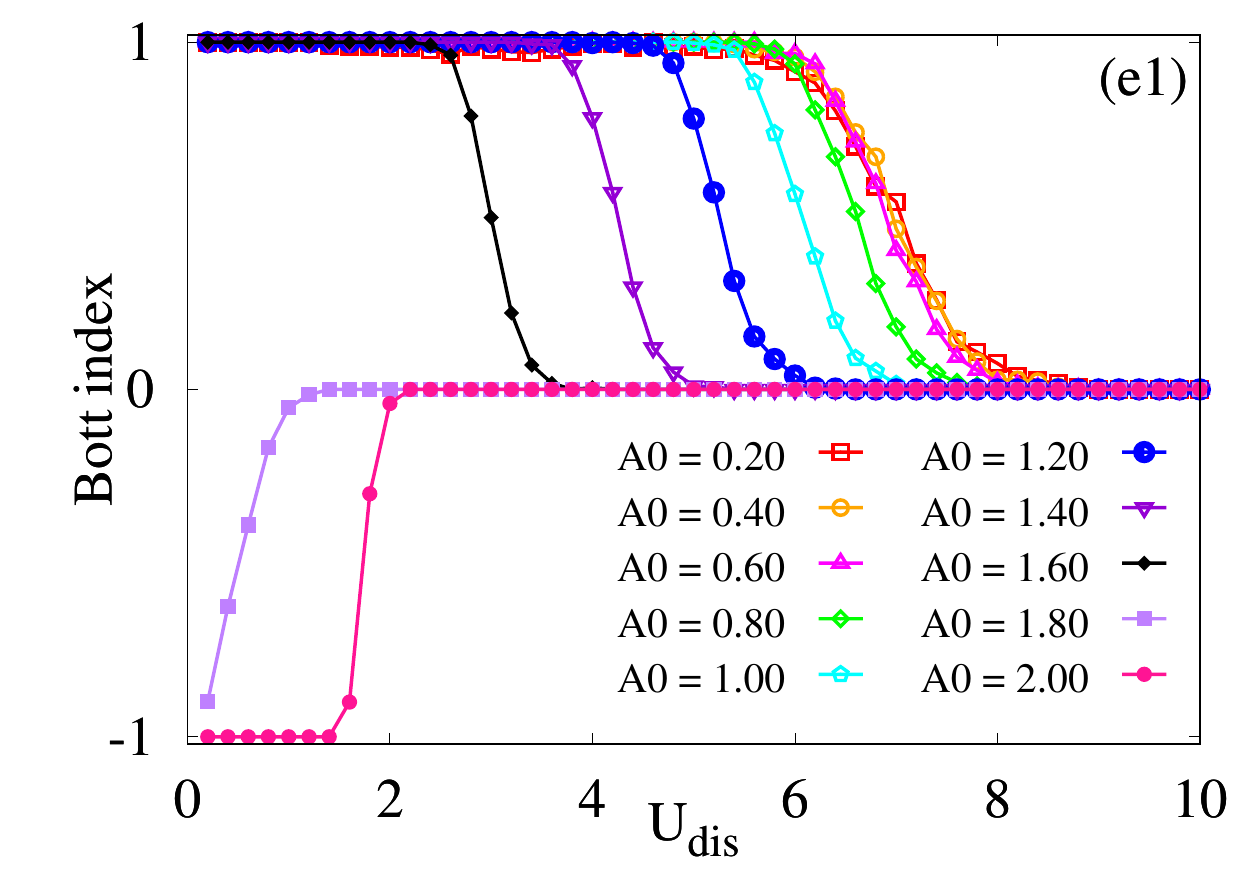}  %
\includegraphics[width=0.33\linewidth]{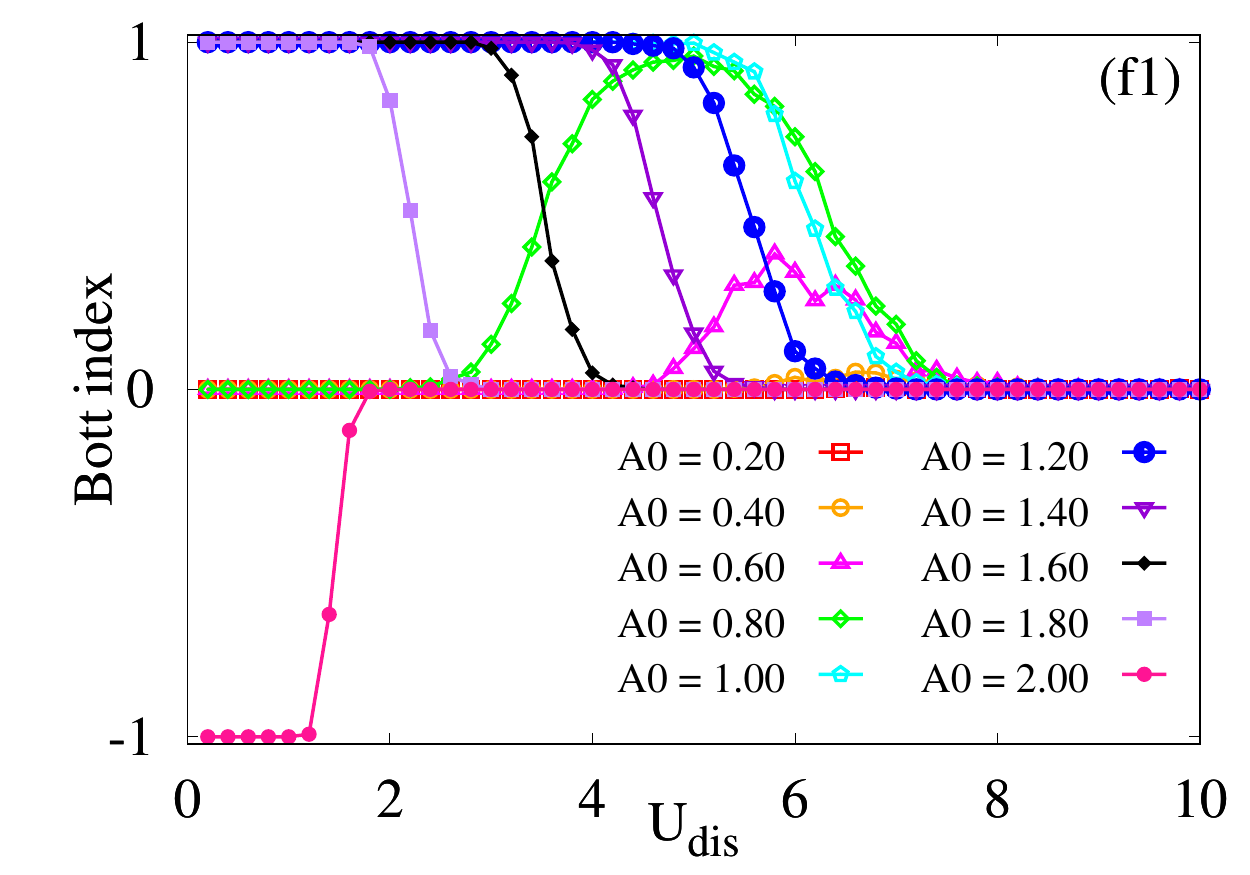}  %
\caption{(Color online) Top panel: The phase diagram in the plane of laser intensity $A_0$ and on-site disorder $U_{dis}$ for both GKM with $t_3 = 0.0$ (a), $t_3 = 0.2$ (b), $t_3 = 0.4$ (c) and DKM with $t_d = 1.5$ (d), $t_d = 2.0$ (e), $t_d = 2.5$ (f). Note the color scale for (c) is different from the others. The phase diagram can serve as a visual guide. The detailed data with Bott index as a function of disorder strength are plotted in the lower panels. Middle panel: Generalized Kane-Mele model with $t_3 = 0.0$ (a1), $t_3 = 0.2$ (b1), $t_3 = 0.4$ (c1) from left to right. Bottom panel: dimerized Kane-Mele model with $t_d = 1.5$ (d1), $t_d = 2.0$ (e1), $t_d = 2.5$ (f1) from left to right. The remaining model parameters are fixed at $t_1=1.0, \lambda=0.3, \Omega=10.0$.
}
\label{fig:bottindex}
\end{figure*}
The position of the ${\bf M}_2$ point is ($-\sqrt{3}\pi/3, \pi/3$), 
\begin{align}
     H_{0} = \begin{pmatrix}
     0 & f_{\bfM_2}  \\
     f_{\bfM_2} & 0
     \end{pmatrix},
\end{align}
with $f_{\bfM_2} = - t_d \mathJ_0(A_0) + 3t_3 \mathJ_0(2A_0)$. The eigenvalues are
\begin{align}
     E_{\pm} &= \pm \left|- t_d \mathJ_0(A_0) + 3t_3 \mathJ_0(2A_0)\right|.
\end{align}
For the generalized Kane-Mele model, we have $t_1 = t_d$, and
the eigenvalues will be 
\begin{align}
    E_{\pm} &= \pm \left|t_1 \mathJ_0(A_0) - 3t_3 \mathJ_0(2A_0)\right|.
\end{align}
For the dimerized Kane-Mele model, the Hamiltonian is independent of the third-neighbor hopping terms $t_3$, and the eigenvalues are 
\begin{align}
     E_{\pm} &= \pm \left|t_d \mathJ_0(A_0)\right|.
\end{align}
The gap for each high symmetry point in the high-frequency limit is summarized in Table \ref{tab:table1}. The gap size at each high symmetry $\bfk$ point is plotted with a dashed line in Fig.\ref{fig:gap}. The exact gap size is plotted with dots, as a comparison. For the Kane-Mele model and the GKM model, the gap calculated using the high-frequency approximation can capture the main feature of the exact results, especially for the gap closing points of $\Gamma, K$, and $M_3$, which correspond to the spin Chern number change. For the DKM Hamiltonian, the high frequency results are in good agreement for both the $\Gamma$ and $K$ points. Apparently, for the $M_3$ points, high-order corrections are needed to explain the gap closing point around $A_0 = 0.8$.
\section{Phase diagram and Bott index for the disordered system}
\label{sec:disorder}
In the top panels of Fig.\ref{fig:bottindex}(a-d), we plot the phase diagram of the GKM model with parameter $t_3 = 0.0, 0.2, 0.4$ and the DKM model with $t_d = 1.5, 2.0, 2.5$. The remaining parameters are fixed at $t_1 = 1.0, \lambda = -0.3, \Omega = 10.0$. The detailed data corresponding to the phase diagram--the Bott index as a function of disorder at different laser intensities--are plotted in the middle panels for GKM with $t_3 = 0.0, 0.2, 0.4$ from left to right and the bottom panels for DKM with $t_d = 1.5, 2.0, 2.5$.
In the clean system limit ($U_{dis} = 0$), the system makes a topological transition as the laser intensity increases, inducing the Dirac points to close and reopen (shown in Fig.\ref{fig:Floquetband}). The inclusion of disorder in the weak disorder region, do not change the original states from topological trivial or non-trivial.  In the strong disorder limit, a topologically trivial (Bott index=0) Anderson insulator appears.

The most interesting phenomena occur for intermediate levels of disorder. Consider Fig.\ref{fig:bottindex} (a), (c), and (f), which represents the Kane-Mele model, the GKM, and the DKM, respectively. Reading the figures horizontally, for fixed disorder strength, as the laser intensity increases, the transition from the topologically non-trivial state to the topologically trivial state occurs in Fig.\ref{fig:bottindex}(a). These results are not easy to explain because the band structure at the starting point with finite disorder strength is not well-defined (momentum is not a good quantum number). As an alternative, one can read the figure vertically, for fixed laser intensity, and study the effect of disorder on the the original Floquet Bloch states. In this way, the starting point is the Floquet-Bloch band structure shown in Fig.\ref{fig:Floquetband} in the first Floquet zone $-\Omega/2 < E_{\bfk} < \Omega/2$. 

Let us focus on Fig.\ref{fig:bottindex}(a) and Fig.\ref{fig:bottindex}(a1) first. We define the critical disorder strength as the point where the Bott index deviates from $1$. A monotonic behavior is observed for $A_0=0.2,0.4, \cdots 1.6$. By inspecting the Floquet band structure for $A_0=0.2,0.4,0.6,0.8$ [Fig.\ref{fig:Floquetband}(a2-a3)], one realizes the band gap at the $M$ point does not change much while the band width is narrowing. This observation explains the results here because for weak laser intensity, the hopping terms are renormalized by a Bessel function $J_{n}(x) < 1$, while the on-site disorder term remains unchanged. Thus, critical disorder will decrease at weak laser intensity.  Further increasing $A_0=1.0,1.2,1.4,1.6$ [Fig.\ref{fig:Floquetband}(a4-a5)], the Floquet-Bloch band structure is significantly changed (the system gap shifts to the $K$ point). In this process, both the bandwidth and system gap decrease, which decreases the critical disorder strength faster.   Finally, at laser intensity $A_0=1.8, 2.0$ [Fig.\ref{fig:Floquetband}(a6)], the bandwidth decreases dramatically while the system gap starts to increase, and the competition between them determines the critical disorder strength. 

Next, we turn to the Bott index as a function of disorder for the generalized Kane-Mele model with $t_3 = 0.4$, shown in Fig.\ref{fig:bottindex}(c1).  First we consider low laser intensity: $A_0 = 0.2,0.3$. As the laser intensity is increased from $A_0=0.2$ to $A_0 = 0.3$, both the bandwidth and the gap at the $M$ point get smaller, which explains why the critical $U_{dis}$ decreases. Around $A_0 = 0.4$, the system gap at the $M$ point closes and reopens. Further increasing the laser intensity to $A_0 = 0.6,0.8$ will increase the gap at the $M$ point, which pushes the critical $U_{dis}$ to larger values.  Further increasing $A_0$ to 1.0 and 1.2, the system gap shifts to the $K$ point (shown in Fig.\ref{fig:Floquetband}(b5)); this pushes the critical disorder to smaller values. The system gap at the $K$ point closes and reopens at $A_0 = 1.4$.  Finally, the minimal gap shifts to the $\Gamma$ point, and further decreases as the laser intensity increases to $A_0 = 2.0$, which explains the critical disorder strength moving to smaller values from $A_0 = 1.8$ to $A_0=2.0$.

Finally, by looking at the data for the dimerized KM model in Fig.\ref{fig:bottindex}(f1), we find a similar story, except differing for $A_0 < 0.8$. We focus our discussion on this region. The starting point here is the topological trivial state with spin Chern number $\mathcal{C} = 0$. For weak laser intensity $A_0 = 0.2, 0.4$, adding disorder does not change the Bott index. The gap is relative large here, and neither weak nor intermediate disorder can close the gap and generate band inversion. Strong disorder, however, will localize all the states.  This idea is confirmed by inspecting the data for $A_0 = 0.6,0.8$. Here the gap at the $M_3$ point gets smaller, and the intermediate disorder strength will close the gap and reopen it, which can be explained by the Born approximation, where the mass is renormalized through disorder. We find the highest values of the data for $A_0 = 0.6, 0.8$ do not reach $1$, which would indicate a topologically non-trivial state. This is explained as a finite size effect because larger system sizes move the Bott index towards 1; more detail is provided as an appendix. 
\begin{figure*}[t]
\includegraphics[width=0.33\linewidth]{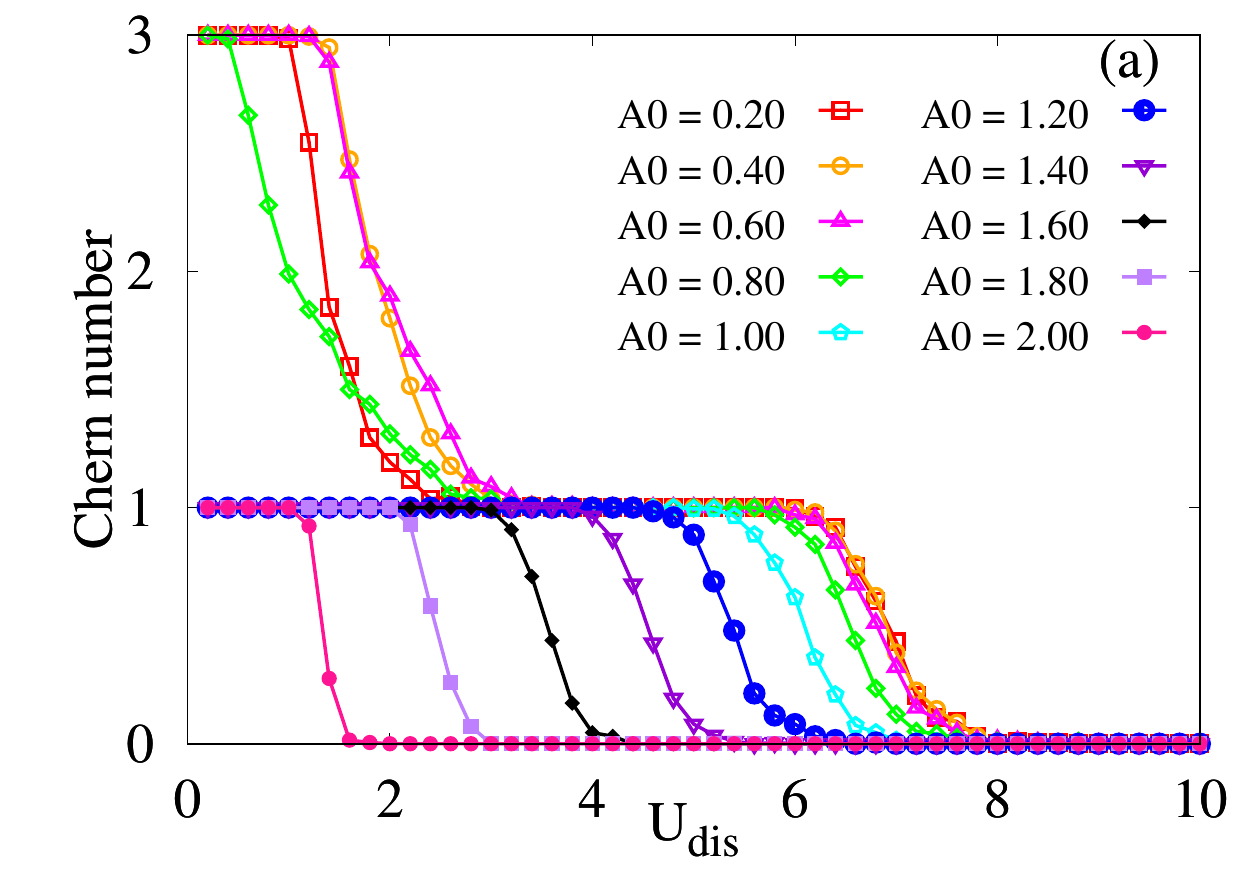}  %
\includegraphics[width=0.33\linewidth]{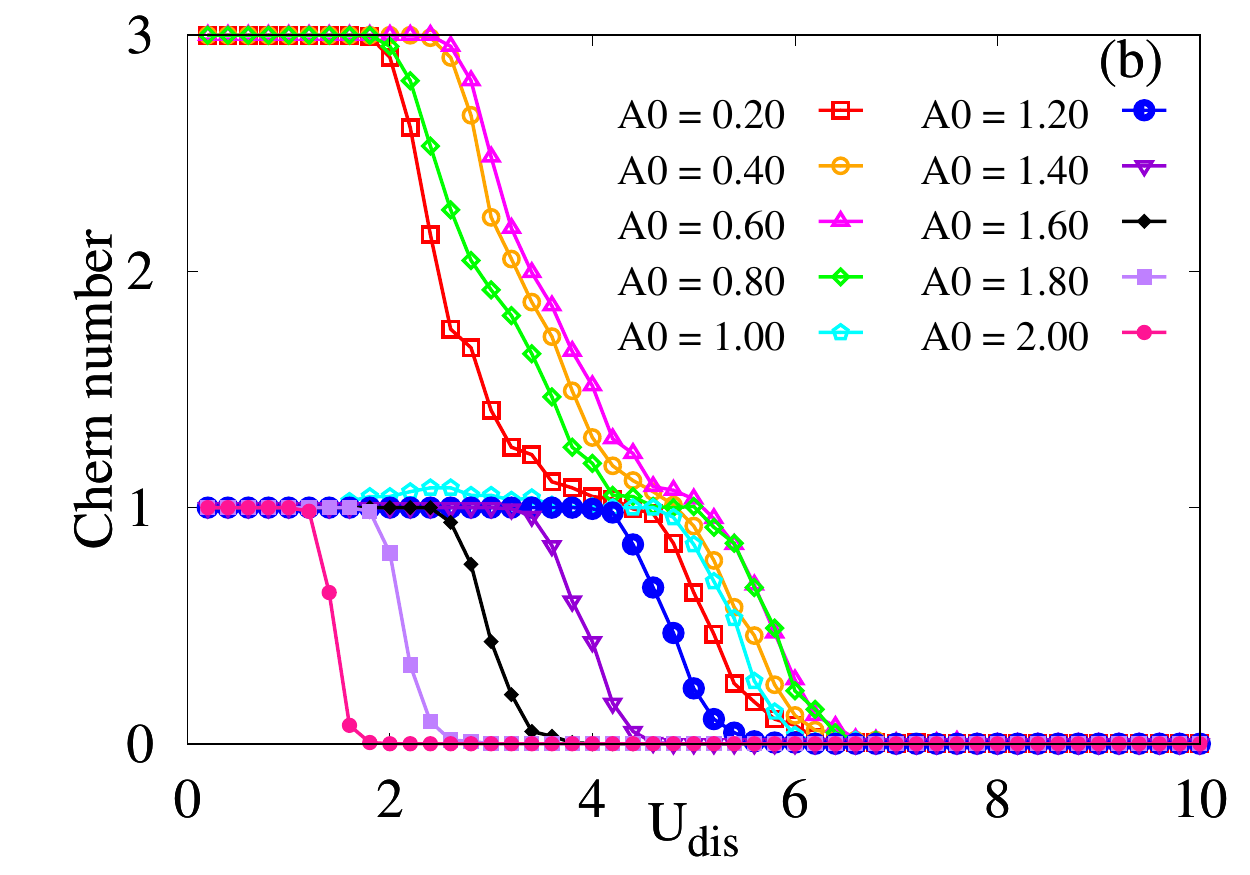}  %
\includegraphics[width=0.33\linewidth]{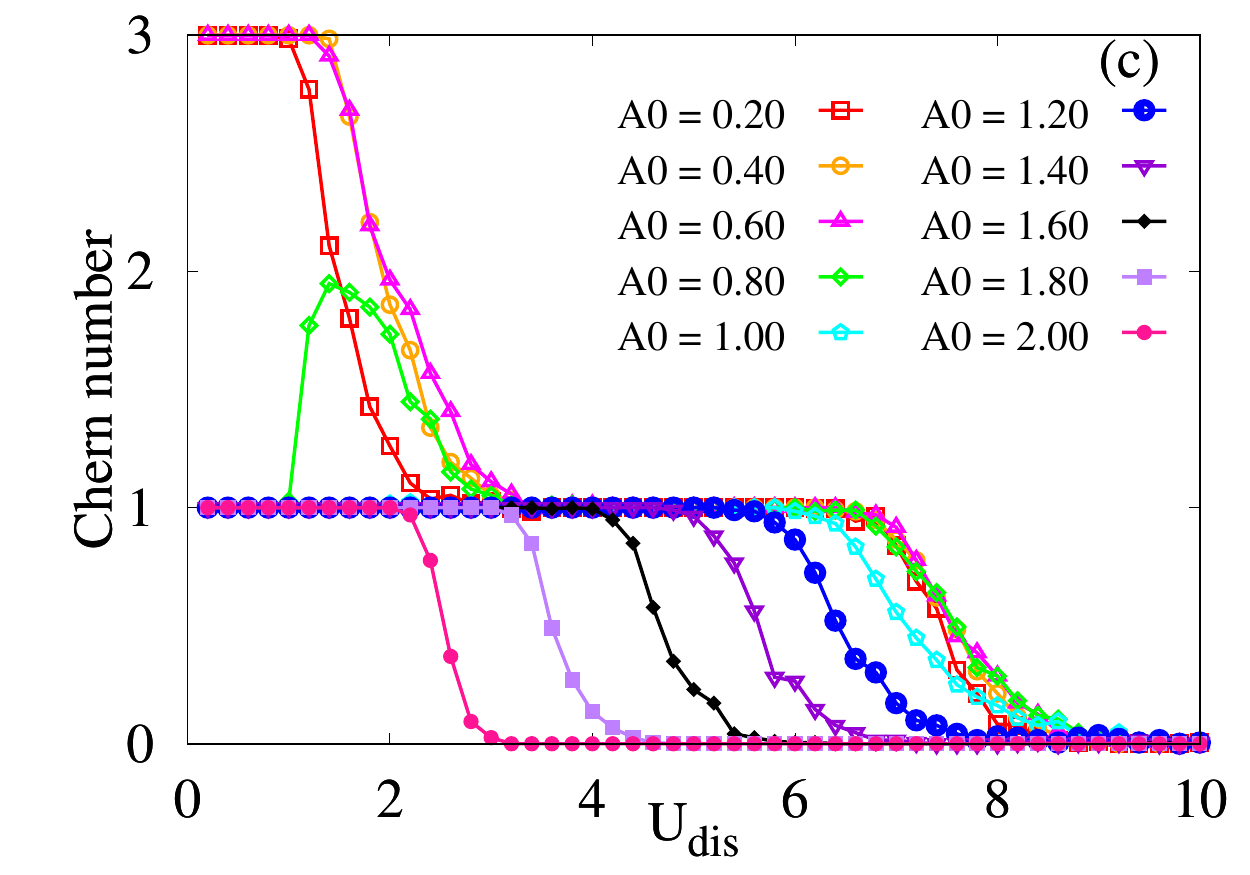}  
\includegraphics[width=0.99\linewidth]{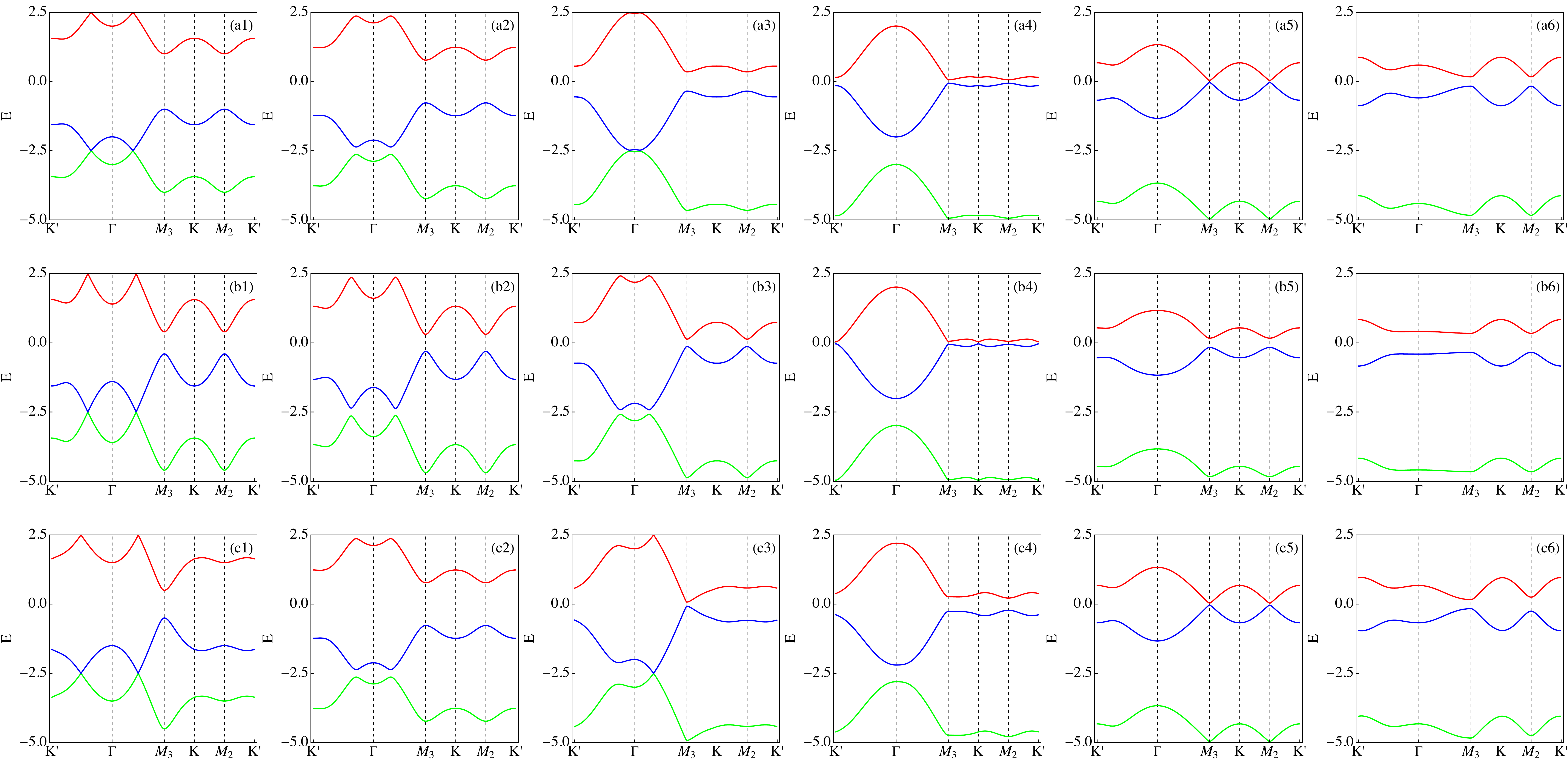}
\caption{(Color online) The Chern number as a function of  on-site disorder $U_{dis}$ for (a) GKM model with $t_3 = 0.0$, (b) GKM model with $t_3 = 0.2$ (c) DKM model with $t_d = 1.5$. The remaining model parameters are fixed at with $t_1=1.0, \lambda=0.3, \Omega=5.0$. The Floquet-Bloch band structure in the clean-limit (absence of disorder) are plotted for different laser intensity $A_0 = 0.0, 0.4, 0.8, 1.2, 1.6, 2.0$ in (a1-a6) for GKM model with $t_3 = 0.0$, (b1-b6) for GKM model with $t_3 = 0.2$ and (c1-c6) for DKM model with $t_d = 1.5$.
}
\label{fig:onresonant}
\end{figure*}

\section{Chern number for disordered system with an on-resonant laser}
In this section, we study the topological invariant as a function of laser intensity and on-site disorder while fixing the laser frequency to be on-resonant ($\hbar\Omega < W$, where $W$ is the bandwidth of equilibrium model Hamiltonian).  In the on-resonant regime, the high-frequency expansion is not expected to be accurate and the system may display a complex evolution as a function of laser parameters.

In the top panels of Fig.\ref{fig:onresonant}, we plot the Chern number as a function of on-site disorder $U_{dis}$ for (a) GKM model with $t_3 = 0.0$ (bandwidth $6t_1$), (b) GKM model with $t_3 = 0.2$ (bandwidth $7.2t_1$) and (c) DKM model with $t_d = 1.5$ (bandwidth $7.0t_1$). The remaining model parameters are fixed at $t_1=1.0, \lambda=0.3, \Omega=5.0$. The laser intensity is varied through $A_0 = 0.2, 0.4, 0.6, \cdots, 2.0$. 
We focus on the clean limit first, increasing the laser intensity from $A_0 = 0.2$ to $A_0 = 2.0$, and note the Chern number will change from $\mathcal{C} = 3$ to $\mathcal{C} = 1$ which is $\Delta\mathcal{C} = 2$. This behavior can be understood by considering the decrease of the laser frequency from infinity to finite on-resonant frequency: At infinite laser frequency, the original equilibrium bandwidth is rescaled by a Bessel function of the first kind. For example, the effective bandwidths $W_{\text{eff}}$ are $6|\mathcal{J}_0(A_0) t_1|$ for the Kane-Mele model, $6|\mathcal{J}_0(A_0) t_1 + 2 \mathcal{J}_0(2A_0) t_3|$ for the GKM model and $2|\mathcal{J}(A_0) (t_d + 2t_1)|$ for the DKM model. The next order correction in the high-frequency limit is a correction to this effective bandwidth.  When the laser frequency is decreased to be equal to the effective bandwidth, the ``top"  of  a ``lower" Floquet copy will touch the ``bottom" of the ``upper" Floquet band at $E = -\Omega/2$. Further decreasing the frequency will generate a quadratic band crossing and a small but finite laser intensity will open a gap between the band crossing, changing the Chern number by $\Delta\mathcal{C} = \pm 2$. 

To further illustrate the picture above, the Floquet-Bloch band structure in the clean-limit (absence of disorder) is plotted for different laser intensities $A_0 = 0.0, 0.4, 0.8, 1.2, 1.6, 2.0$ in (a1-a6) for the KM model with $t_3 = 0.0$, (b1-b6) for the GKM model with $t_3 = 0.2$ and (c1-c6) for the DKM model with $t_d = 1.5$. 
The quasi-energy bands are plotted from $-\Omega$ to $\Omega/2$ which includes the copy in the Floquet zone $-\Omega/2 < \epsilon < \Omega/2$ and half of the lower copy $-\Omega < \epsilon < -\Omega/2$ to show the band crossing point at $\Gamma$. We focus on the behavior of the Chern number with the laser intensity $A_0 = 0.8$. In the KM model, the Floquet-Bloch band structure is shown in Fig.\ref{fig:onresonant}(a3). The system gap is situated very close to the $\Gamma$ point and is small compared to the system gap at the $M_3$ point. In this way, a small amount of disorder will close the gap around the $|\Gamma|$ point first (changing the Chern number by 2), and then close the gap at the $M_3$ point, changing the Chern number by 1. The magnitude $\mathcal{C}=1$ is the result of the gap differences at energy $E=-\Omega/2$ and $E=0.0$. This picture is confirmed by comparing the data for the GKM model with $t_3 = 0.2$ [shown in Fig.\ref{fig:onresonant}(b) and (b3)]. Since the original bandwidth of the model is larger than the bandwidth of KM model, the gap formed at $E=-\Omega/2$ is larger.  This may generate the larger critical disorder to change the Chern number by $\pm 2$. Secondly, the energy gap difference at energy at $E=-\Omega/2$ and $E=0.0$ is relatively smaller, which induces the smaller magnitude of $\mathcal{C}=1$. 

For the DKM model with $A_0=0.8$, the Chern number changes from $1$ to $2$ with small disorder strength and comes back to $1$ as the disorder increases.  By inspecting the Floquet-Bloch band structure in Fig.\ref{fig:onresonant}(c3), we realize there is a linear crossing between the $\Gamma$ and $M_3$ points. A small amount of disorder can induce an effective mass which generate a band inversion and a Chern number change $\pm 1$. Further increasing the intensity will close the gap and bring one back to $\mathcal{C} = 1$. Continuing to increase the disorder will induce the transition from $1$ to $0$, which is determined by the energy gap at $E=0.0$.

\section{Conclusion}
\label{sec:conclusion}
In this paper we theoretically studied the topological properties of the generalized Kane-Mele (GKM) model with third-neighbor hopping $t_3$ and the dimerized Kane-Mele (DKM) model with dimerized hopping $t_d$ along the vertical direction [along $\delta_3$ in Fig.\ref{fig:honeycomb}(a)] under illumination by a circularly polarized monochromatic laser field.  In the absence of the laser, the GKM model has a critical value of $t_3 = 1/3$, where topological trivial and non-trivial states occur for values larger and smaller than the critical $t_3$, respectively. The DKM model has critical $t_d = 2.0$ where topological trivial and non-trivial states occur for values larger and smaller than the critical $t_d$, respectively. 

To include both topologically trivial and non-trivial states as starting points, we chose $t_3 = 0.0, 0.2, 0.4$ for the GKM model and $t_d = 1.5, 2.0, 2.5$ for the DKM model. Their complicated phase structures were studied numerically, both in the high-frequency off-resonant case and the low-frequency resonant case.  The topological transitions are explained using the Floquet-Bloch band structure, where we find the laser will close and reopen Dirac points, inducing a Chern number change $\Delta\mathcal{C} = \pm 1$ for each Dirac point. Further, we found the laser can shift the system gap between different high symmetry points.  For example, the minimal gap may shift from an $M$ point to a $K$ point in the Kane-Mele model [shown in Fig.\ref{fig:Floquetband}(a1-a6)] or even shift to some point without high symmetry for the DKM model [shown in Fig.\ref{fig:Floquetband}(c1-c6)]. The band structure, and the system gap at high symmetry points, is explained using the low-energy Hamiltonian based on a high frequency expansion for the off-resonant case.

Finally, we study the effect of on-site disorder in the GKM and DKM model under a periodic laser drive (Floquet system). Topological states are sustained with weak disorder, and destroyed by strong disorder, similar to the case in equilibrium. In addition, weak disorder may even generate a topologically trivial state from a non-trivial one providing a level of material control through the interplay of disorder and a periodic drive.  Compared to the more heavily studied Kane-Mele model with disorder, the minimal gap evolution through the Brillouin zone for teh GKM and DKM models presents new phenomenology for disordered Floquet systems.

\section*{appendix: Finite size effect}
The finite size effect on the non-quantized
region of the Bott index where the Floquet-Anderson topological transition occurs is studied here. In Fig.\ref{fig:bottindex}(c1) there exists a plateau around the Bott index $1$ with $A_0 = 0.2,0.3$ and (f1) the Bott index does not reach 1 with $A_0 = 0.6,0.8$. Here we studied the two cases with different size to check what the finite size effect is.

\begin{figure}[t]
\includegraphics[width=0.49\linewidth]{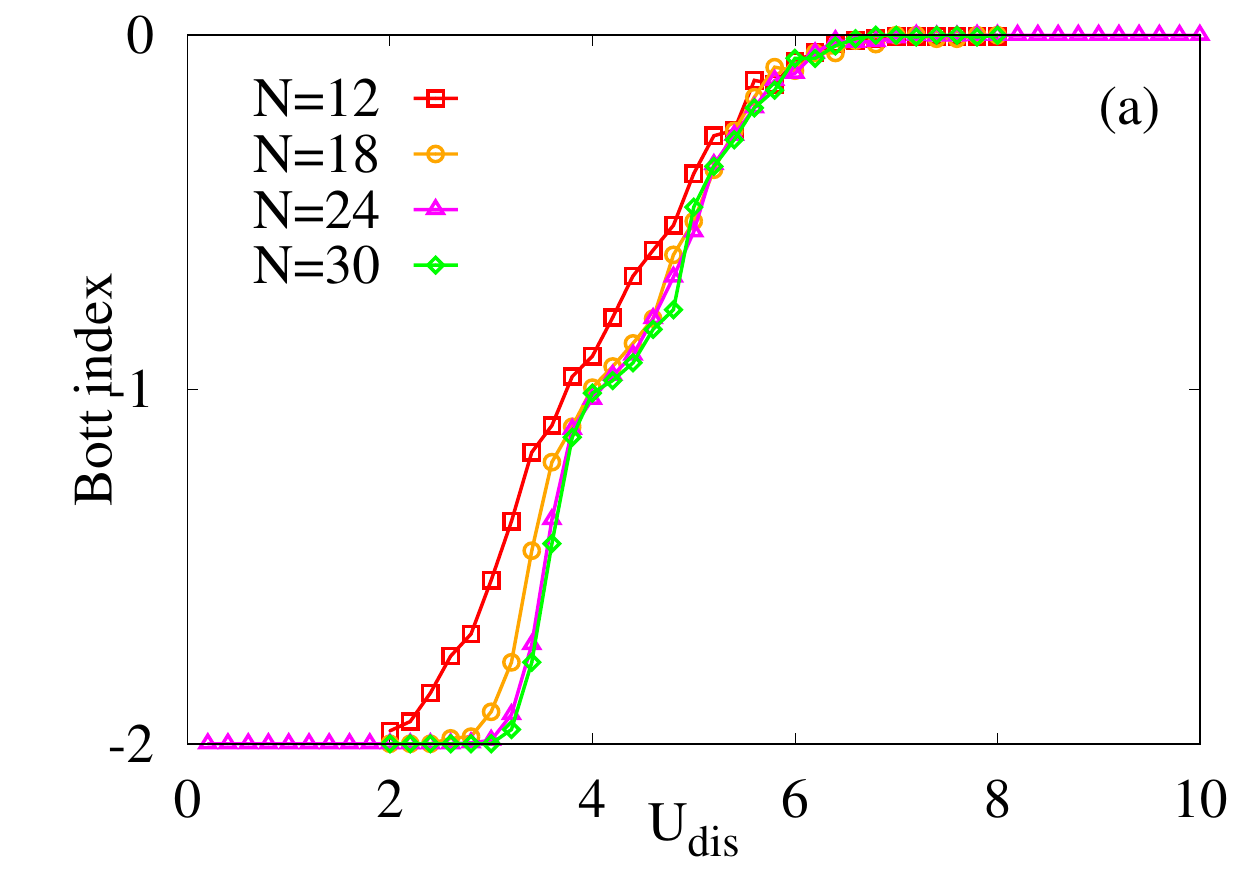}  %
\includegraphics[width=0.49\linewidth]{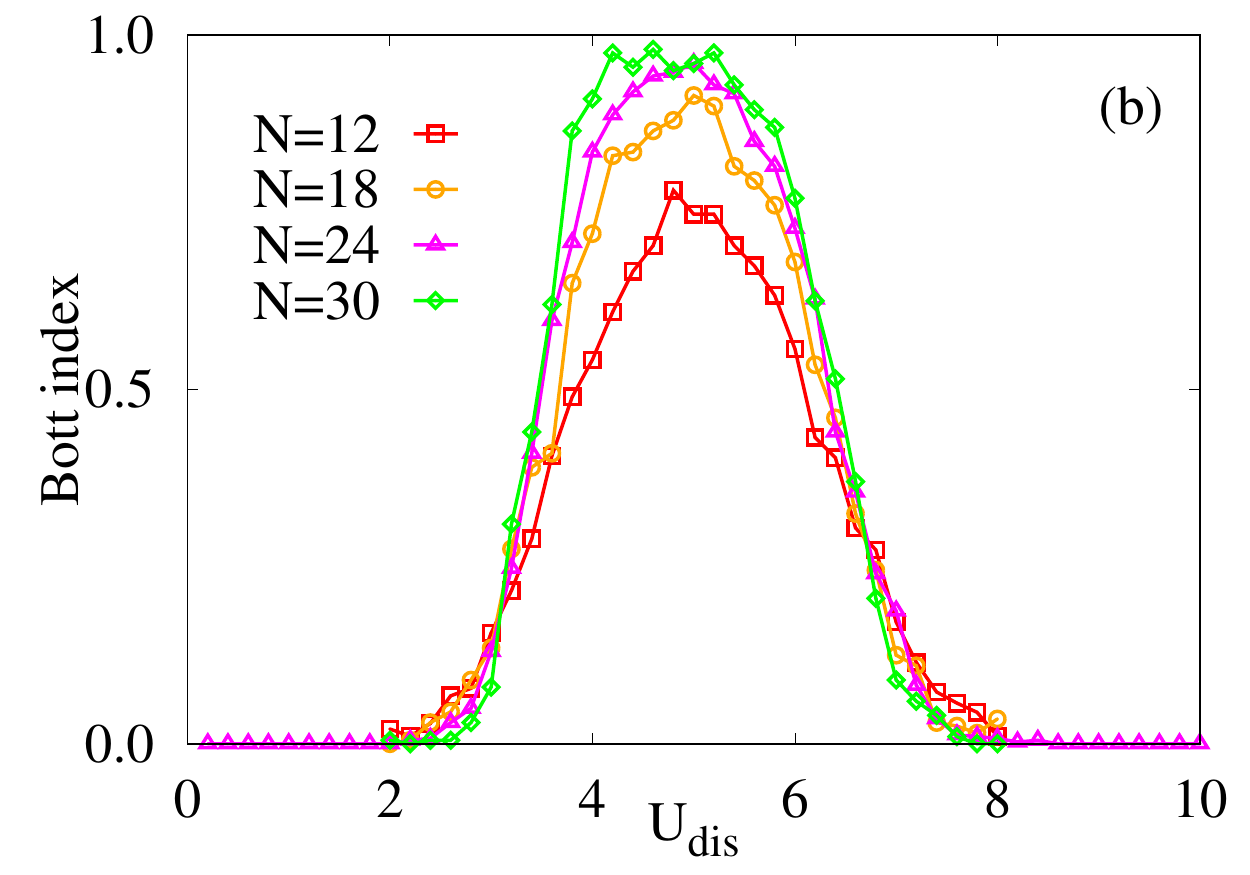}  %
\caption{(Color online) (a) The Bott index as a function of  on-site disorder $U_{dis}$ for the GKM with $t_3 = 0.4$ and laser intensity $A_0 = 0.2$. (b) The Bott index as a function of on-site disorder $U_{dis}$ for the DKM model with $t_d = 2.5$ and laser intensity $A_0 = 0.8$. The remaining model parameters are fixed at $t_1=1.0, \lambda=0.3, \Omega=10.0$. Different cluster sizes are chosen to illustrate the finite size effect. The total number of lattice sites are $N \times N \times 2$ where the 2 comes from the number of atoms in one unit cell.
}
\label{fig:finitesize}
\end{figure}

In Fig.\ref{fig:finitesize}, we plot the disorder- averaged Bott index as a function of disorder for different system sizes. It is clear that with increasing system size, the non-quantized region of the Bott index becomes sharper, which is consistent with previous studies.\cite{Titum:prl15,Hung:prb16} Further, in Fig.\ref{fig:finitesize}(b), we realize there will be a quantized area with increasing cluster size.

\section*{Acknowledgements} 
We acknowledge helpful discussions with Hsiang-Hsuan Hung, Yang Ge, Quansheng Wu, and Yingyue Boretz. We gratefully acknowledge funding from Army Research Office Grant No. W911NF-14-1-0579, NSF Grant No. DMR-1507621, and NSF Materials Research Science and Engineering Center Grant No. DMR-1720595.
\bibliography{ftai.bib}
\end{document}